\documentclass[aps,prb,twocolumn,showpacs,superscriptaddress,groupedaddress]{revtex4}  
\usepackage{graphicx}  
\usepackage{dcolumn}   
\usepackage{bm}        
\usepackage{amssymb}   
\usepackage{color}
\usepackage{physics}
\usepackage[colorlinks = true,
            linkcolor = blue,
            urlcolor  = blue,
            citecolor = blue,
            anchorcolor = blue]{hyperref}
\usepackage[all]{hypcap}

\usepackage[applemac]{inputenc}
\usepackage{amsmath}
\usepackage{mathptmx}

\newcommand{\refsub}[2][{}]{\hyperref[#2]{\ref{#2}#1}} 

\newcommand{\sdo}{SrDy$_2$O$_4$}
\newcommand{\sho}{SrHo$_2$O$_4$}
\newcommand{\seo}{SrEr$_2$O$_4$}

\newcommand{\uud}{$\uparrow \uparrow \downarrow$}

\newcommand{\uudd}{$\uparrow \uparrow \downarrow \downarrow$}
\newcommand{\udud}{$\uparrow \downarrow \uparrow \downarrow$}
\newcommand{\uuuu}{$\uparrow \uparrow \uparrow \uparrow$}

\begin{document}

\title{Absence of long range order in the frustrated magnet \sdo\ due to trapped defects from a dimensionality crossover}

\author{N. Gauthier}  
\email{nicolas.gauthier4@gmail.com}
\affiliation{Laboratory for Scientific Developments and Novel Materials, Paul Scherrer Institut, 5232 Villigen, Switzerland}

\author{A. Fennell} 
\affiliation{Laboratory for Neutron Scattering and Imaging, Paul Scherrer Institut, 5232 Villigen, Switzerland}

\author{B. Pr\'evost}
\affiliation{D\'epartement de Physique \& Regroupement Qu\'eb\'ecois sur les Mat\'eriaux de Pointe (RQMP), Universit\'e de Montr\'eal, Montr\'eal, Qu\'ebec H3C 3J7, Canada}

\author{A.-C. Uldry}
\affiliation{Condensed Matter Theory Group, Paul Scherrer Institut, 5232 Villigen, Switzerland}

\author{B. Delley}
\affiliation{Condensed Matter Theory Group, Paul Scherrer Institut, 5232 Villigen, Switzerland}

\author{R. Sibille}
\affiliation{Laboratory for Neutron Scattering and Imaging, Paul Scherrer Institut, 5232 Villigen, Switzerland}

\author{A. D\'esilets-Benoit}
\affiliation{D\'epartement de Physique \& Regroupement Qu\'eb\'ecois sur les Mat\'eriaux de Pointe (RQMP), Universit\'e de Montr\'eal, Montr\'eal, Qu\'ebec H3C 3J7, Canada}

\author{H.A. Dabkowska} 
\affiliation{Brockhouse Institute for Materials Research, Hamilton, Ontario L8S 4L8, Canada}

\author{G.J. Nilsen}
\affiliation{Institut Laue-Langevin, CS 20156, F-38042 Grenoble Cedex 9, France}

\author{L.-P. Regnault}
\affiliation{Institut Laue-Langevin, CS 20156, F-38042 Grenoble Cedex 9, France}

\author{J.S. White}
\affiliation{Laboratory for Neutron Scattering and Imaging, Paul Scherrer Institut, 5232 Villigen, Switzerland}

\author{C. Niedermayer}
\affiliation{Laboratory for Neutron Scattering and Imaging, Paul Scherrer Institut, 5232 Villigen, Switzerland}

\author{V. Pomjakushin}
\affiliation{Laboratory for Neutron Scattering and Imaging, Paul Scherrer Institut, 5232 Villigen, Switzerland}

\author{A.D. Bianchi} 
\affiliation{D\'epartement de Physique \& Regroupement Qu\'eb\'ecois sur les Mat\'eriaux de Pointe (RQMP), Universit\'e de Montr\'eal, Montr\'eal, Qu\'ebec H3C 3J7, Canada}

\author{M. Kenzelmann}
\email{michel.kenzelmann@psi.ch}
\affiliation{Laboratory for Scientific Developments and Novel Materials, Paul Scherrer Institut, 5232 Villigen, Switzerland}
\vskip 0.25cm
 
\date{\today}

\begin{abstract}
Magnetic frustration and low dimensionality can prevent long range magnetic order and lead to exotic correlated ground states. \sdo\ consists of magnetic Dy$^{3+}$ ions forming magnetically frustrated zig-zag chains along the $c$-axis and shows no long range order to temperatures as low as $T=60$~mK. We carried out neutron scattering and AC magnetic susceptibility measurements using powder and single crystals of \sdo. Diffuse neutron scattering indicates strong one-dimensional (1D) magnetic correlations along the chain direction that can be qualitatively accounted for by the axial next-nearest neighbour Ising (ANNNI) model with nearest-neighbor and next-nearest-neighbor exchange $J_1=0.3$~meV and $J_2=0.2$~meV, respectively. Three-dimensional (3D) correlations become important below $T^*\approx0.7$~K. At $T=60$~mK, the short range correlations are characterized by a putative propagation vector $\textbf{k}_{1/2}=(0,\frac{1}{2},\frac{1}{2})$. We argue that the absence of long range order arises from the presence of slowly decaying 1D domain walls that are trapped due to 3D correlations. This stabilizes a low-temperature phase without long range magnetic order, but with well-ordered chain segments separated by slowly-moving domain walls.

\end{abstract}


\maketitle

\section{Introduction}
While most magnetic systems feature long range ordered ground states, magnetic frustration can preclude magnetic order to the lowest temperatures and instead promote strongly correlated fluctuating phases. Absence of magnetic order due to magnetic frustration was observed in a number of rare-earth pyrochlore oxides exhibiting topological magnetic phenomena. For example, Dy$_2$Ti$_2$O$_7$ and Ho$_2$Ti$_2$O$_7$ feature spin ice states at low temperature where the spin disorder is analogous to the proton position disorder in water ice.\cite{Harris1997,Ramirez1999} Other examples include the spin glass state of Y$_2$Mo$_2$O$_7$ in absence of structural disorder  \cite{Gingras1996} and the magnetoelastic spin liquid state of Tb$_2$Ti$_2$O$_7$.\cite{Fennell2014a} 

Magnetic interactions which dominate along mainly one or two spatial dimensions can also weaken the tendency towards magnetic order. Purely one-dimensional (1D) systems can not order at finite temperature.\cite{Mermin1966} The magnetic order in such materials is generally stabilized due to weak interchain interactions leading to two- or three-dimensional (3D) order at temperatures much lower than the Curie-Weiss temperature. The simultaneous presence of frustration and low dimensionality can lead to novel phenomena and complex phase diagrams, of which the zig-zag chain is a case in point. The zig-zag chain is described by nearest-neighbour exchange $J_1$ and next-nearest-neighbour exchange $J_2$, where $J_2$ acts as the source of frustration. For Ising spins, different ground states are stabilized depending on the size of the antiferromagnetic $J_2>0$. For ratio $\frac{J_2}{|J_1|}<0.5$ and an antiferromagnetic $J_1$ a simple N\'eel state \udud\ is realized while a ferromagnetic $J_1$ leads to a ferromagnetic state \uuuu. For ratio $\frac{J_2}{|J_1|}>0.5$ a double N\'eel state \uudd\ is realized, regardless of the sign of $J_1$.\cite{Morita1972} These states are separated by a critical point at $\frac{J_2}{|J_1|}=0.5$ where the degree of frustration is strongest. For antiferromagnetic $J_1$ and $J_2$, a magnetic field stabilizes an intermediate \uud\ state, associated with a magnetization plateau at 1/3 of the full saturation~$M_s$.\cite{Oguchi1965,Morita1972,Rujan1983}

\begin{figure}[!htb]
\includegraphics[scale=1.1]{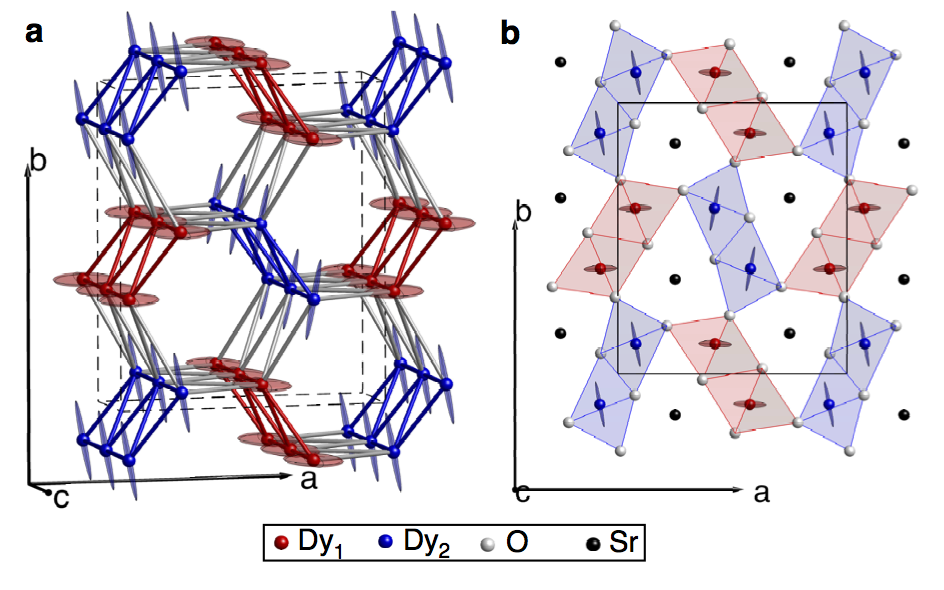}
\caption{(Color online) (a) Magnetic lattice of \sdo\ showing both inequivalent sites in red and blue with their respective $g$-factors represented by ellipsoids. (b) \sdo\ structure in the $ab$ plane showing the oxygen octahedra surrounding the Dy$^{3+}$ ions.}
\label{fig:structure}
\end{figure}

The magnetism in many members of the \textit{A}\textit{R}$_2$O$_4$ (\textit{A}~=~Ba, Sr and \textit{R}~=~Nd, Gd, Tb, Dy, Ho, Er, Tm and Yb) family is qualitatively well described by the zig-zag chain model, such as \sdo\ and \sho.\cite{Fennell2014} These compounds are frustrated magnets with various interesting properties: they feature spin-liquid-like ground-states,\cite{Aczel2015} magnetic phases with multiple coexisting order parameters,\cite{Young2013,Wen2015,Hayes2011} low-dimensional correlations,\cite{Fennell2014} magnetization plateaus \cite{J.Hayes2012} and magnetic field induced order.\cite{Cheffings2013,Bidaud2016,Young2014,Quintero-Castro2012} 

The magnetic rare earth ions form a distorted honeycomb lattice in the $ab$ plane and zig-zag chains along the $c$-axis. There are two inequivalent rare earth sites, surrounded by distorted oxygen octahedra (Fig.~\ref{fig:structure}). These two inequivalent sites form two inequivalent zig-zag spin chains. This magnetic topology has also been observed in $\beta$-CaCr$_2$O$_4$ \cite{Damay2010} and CaV$_2$O$_4$ \cite{Kikuchi2001} and low dimensionality correlations are a common feature in these compounds. 

\sdo\ is particularly interesting because no long range order has been observed in zero field to the lowest temperature surveyed. Previous powder neutron scattering results show the presence of 1D correlations persisting down to $T=60$~mK,\cite{Fennell2014} providing evidence that magnetic interactions along the chain direction dominate the cooperative magnetic properties. Furthermore, this system exhibits a magnetization plateau at $\frac{1}{3}M_s$ for field along the $b$-axis,\cite{J.Hayes2012} as expected in the zig-zag chain for antiferromagnetic $J_1$ and $J_2$. The field-induced phase related to this plateau has been mapped by specific heat \cite{Cheffings2013} and more recently by ultrasound measurements.\cite{Bidaud2016}

We studied the crystal structure of \sdo\ by neutron diffraction and revisited the crystal electric field level scheme, which we discuss in section \ref{sec:structure}. We present in section \ref{sec:ZF} a detailed single-crystal neutron scattering study of the diffuse scattering at $T=60$~mK. We also present AC susceptibility measurements in section \ref{sec:AC}. The results are discussed in section IV in the context of the axial next-nearest neighbour Ising (ANNNI) model and the domain walls of its ground state correlations.

\section{Experimental details}
\sdo\ powder was prepared by solid state reaction using high purity starting materials, similar to the method described by Balakrishnan \textit{et al.}\cite{Balakrishnan2009} Stoichiometric mixtures of Dy$_2$O$_3$ (99.995$\%$) and SrCO$_3$ (99.994$\%$), with a 1$\%$ surplus of carbonate, both dried at 600$^\circ$C and weighted in glovebox, were mixed in a ball mill and pressed into a rod. The rods were heated in air at 1400$^\circ$C in an alumina crucible for a total of three days with two intermediate steps of grinding and pressing. Single crystals were grown using an optical floating zone furnace equipped with four xenon arc lamps. Polycrystalline rods of about 8 cm were pressed, sintered in air at 1500$^\circ$C and used as seeds and feeds, with a growth rate of 10 mm/hour in an ultra pure argon environment. 

Powder neutron diffraction was carried out on the HRPT diffractometer, SINQ at the Paul Scherrer Institut. The \sdo\ powder was loaded in a double-walled copper can ($R_\text{inner}=4.5$~mm, $R_\text{outer}=5.0$~mm) with 10 bar of helium exchange gas for better thermalization in the dilution refrigerator. The patterns were collected using neutrons with a wavelength $\lambda=1.155~\rm{\AA}$ and $1.886~\rm{\AA}$ from temperature $T=0.1$~K up to 100~K. The D7 neutron diffractometer at ILL was used to probe the diffuse scattering of \sdo\ with XYZ polarisation analysis.\cite{Poole2013} Two samples of coaligned single crystals with a total mass of about 0.1 g were measured in the $(h0l)$ and $(0kl)$ reciprocal planes. The samples were glued on a silicon plate with araldite and CYTOP, and mounted on a copper holder. The measurements were carried out in a dilution refrigerator using a neutron wavelength $\lambda=3.1$~\AA. Prior to the data collection, vanadium, quartz and cadmium were measured to correct the data for normalization of the detector efficiency, the flipping ratio and the background, respectively. Measurements taken at $T=40$~K were used to subtract the background scattering due to both the sample holder and the sample environment from the low temperature scattering. Diffuse scattering was also measured on TASP and RITA-II triple-axis spectrometers at SINQ as well as on the IN22 triple-axis spectrometer at ILL. For TASP, the sample was oriented in the $(0kl)$ plane and made of coaligned 300~$\mu$m thick single crystals covering the full surface of a $2 \times 3$~cm$^2$ copper plate, for a total mass of $\sim$1~g. The crystals were fixed using CYTOP. The sample was inserted in a dilution refrigerator and the measurements were carried out at fixed analyzer energy $E_f=3.5$~meV with a beryllium filter. For RITA-II, a sample of coaligned single crystals mounted on an aluminum plate, oriented in the $(h0l)$ scattering plane and inserted in a dilution refrigerator was measured with $E_f=13.7$~meV. IN22 was used in polarized mode with the CryoPad device at fixed $E_f= 14.68$~meV and 30.55~meV. The sample consisted of several coaligned crystals with a total mass of 120 mg mounted on a silicon plate in a copper mount. Measurements were performed in an orange cryostat at $T=1.4$~K.

AC susceptibility measurements were performed in a Quantum Design MPMS equipped with an iHelium option. The sample with dimensions $0.9 \times 1.9 \times 0.4$~mm$^3$ was measured in zero field for an excitation amplitude of 1~Oe along the $b$-axis from $T=0.6$~K up to 1.8~K. Due to the large magnetic moments in \sdo, the measurements had to be corrected for demagnetization.\cite{KMatsuhiraetal2001} The demagnetization factor $N=0.128$ was calculated from the sample dimensions using the equations for a rectangular prism.\cite{Aharoni1998} In our case, this correction did not qualitatively change the AC susceptibility of \sdo, although this was the case previously in other measurements.\cite{Quilliam2011}

\section{Experimental results}

\subsection{Crystal structure and crystal field excitations}
\label{sec:structure}
\begin{figure}[!htb]
\includegraphics[scale=0.55]{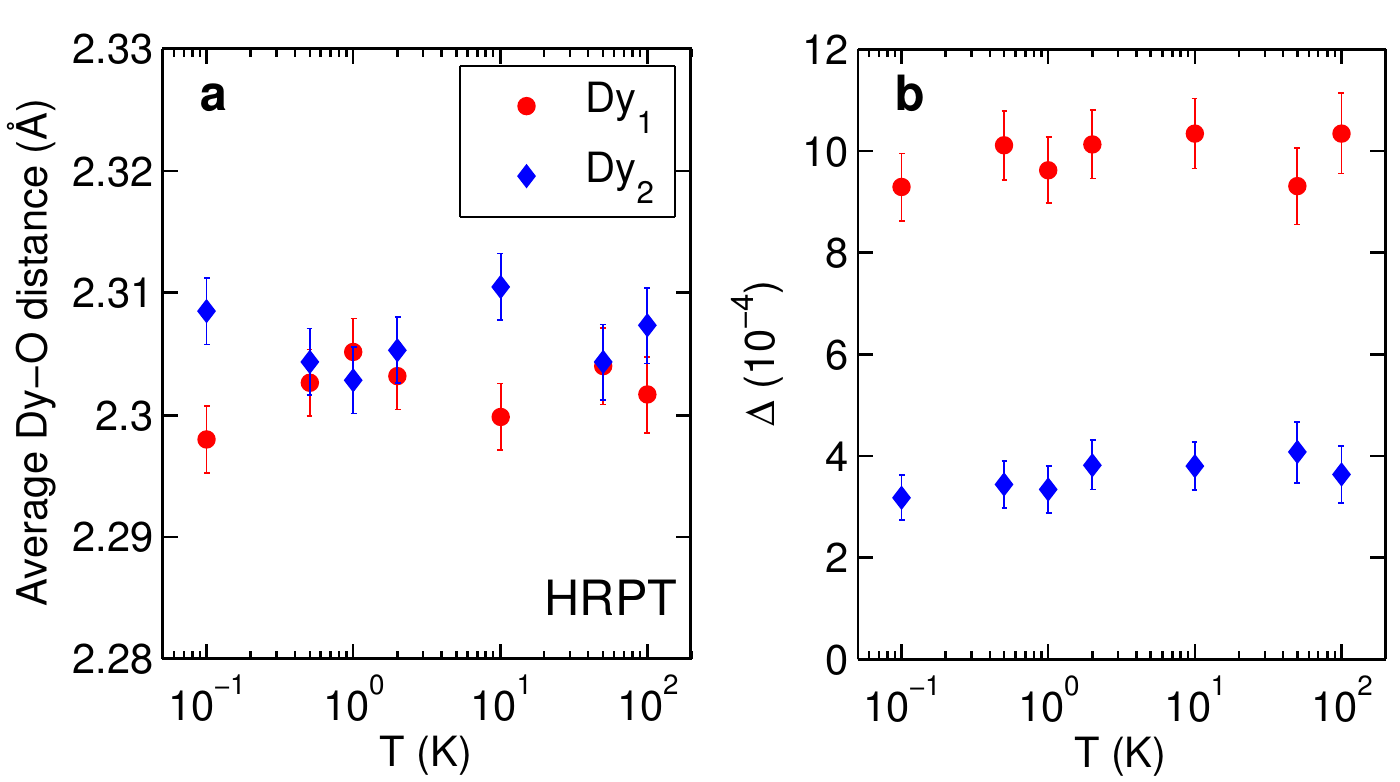}
\caption{(Color online) Temperature dependence of (a) the average Dy-O bond distance and (b) the distortion $\Delta$ (Eq. \ref{distort}) for the oxygen octahedron of each inequivalent Dy$^{3+}$ site, based on combined refinement of powder diffraction pattern recorded with $\lambda=1.155~\rm{\AA}$ and $1.886~\rm{\AA}$ at HRPT.}
\label{HRPT/distortion}
\end{figure}

Powder neutron diffraction reveal no significant structural changes between $T=0.1$~K and 100~K. Diffraction patterns were measured using both a wavelength of $\lambda=1.155~\rm{\AA}$ and $1.886~\rm{\AA}$ for each temperature and the structure was determined by combined Rietveld refinement using \textsc{FullProf}.\cite{Fullprof} The broad magnetic diffuse scattering at low temperature was treated as background to refine the nuclear structure. The most important structural change that may affect the magnetism is related to the oxygen octahedra surrounding the Dy$^{3+}$ ions, which determine the crystal fields that define the single ion anisotropy. The average octahedral distortion $\Delta$ is characterized by
\begin{equation}
\Delta = \frac{1}{6} \sum_{n=1}^6 \left( \frac{d_n-\langle d \rangle}{\langle d \rangle} \right)^2 
\label{distort}
\end{equation}
where $d_n$ is the distance between the Dy$^{3+}$ ions and the \textit{n}th oxygen atom and $\langle d \rangle$ is the average value of $d_n$. Figure \ref{HRPT/distortion} shows that the average Dy-O bond lengths and the distortion $\Delta$ for both inequivalent sites are temperature independent within the accuracy of our measurements. From our results, we can exclude a structural phase transition in \sdo\ between $T=0.1$~K and 100~K. We point out, however, that more precise measurements on SrTm$_2$O$_4$ and SrTb$_2$O$_4$ show a clear temperature dependence of $\Delta$ and a similar distortion in both compounds.\cite{Li2015,Li2014a}

\begin{table}
\begin{tabular}{c|ccc}
  \hline       \hline
  $T$ (K) & \multicolumn{3}{c}{0.10(2)} \\
   Space group & \multicolumn{3}{c}{$Pnam$}\\ 
  $a$ ($\rm{\AA}$) & \multicolumn{3}{c}{10.0773(3)} \\
  $b$ ($\rm{\AA}$) & \multicolumn{3}{c}{11.9124(4)} \\
  $c$ ($\rm{\AA}$)  & \multicolumn{3}{c}{3.4240(1)} \\
   $\alpha$, $\beta$, $\gamma$ ($^\circ$)  & \multicolumn{3}{c}{90, 90 ,90} \\
   \hline
$\lambda= 1.155 \rm{\AA}$ & \multicolumn{3}{c}{ } \\
  \ $R_p$, $R_{wp}$, $R_{exp}$, $\chi^2$ & \multicolumn{3}{c}{13.4, 13.9, 9.85, 2.002} \\
  \hline
$\lambda=1.886 \rm{\AA}$ & \multicolumn{3}{c}{ } \\
      \ $R_p$, $R_{wp}$, $R_{exp}$, $\chi^2$ & \multicolumn{3}{c}{11.2, 9.86, 6.77, 2.122} \\
        \hline
      Global $\chi^2$ & \multicolumn{3}{c}{2.06}\\
   \hline     \hline
    Atom & $x$ & $y$ & $z$  \\
  \hline 
Sr & 0.75  & 0.65 & 0.25 \\
Dy$_1$ & ~~~0.4239(4)~~~ & ~~~0.1109(2)~~~ & 0.25 \\
Dy$_2$ & 0.4210(4) & 0.6114(2) & 0.25 \\
O$_1$ & 0.2150(8) & 0.1771(6) & 0.25 \\
O$_2$ & 0.1297(6) & 0.4797(7) & 0.25 \\
O$_3$ & 0.5113(8) & 0.7838(6) & 0.25 \\
O$_4$ & 0.4248(9) & 0.4221(5) & 0.25 \\
    \hline       \hline   
\end{tabular}
\caption{Determined structure of \sdo\ for combined refinement of powder neutron data with $\lambda=1.155 \rm{\AA}$ and $1.886 \rm{\AA}$ at HRPT.}
\label{tablehrpt}
\end{table}

\begin{figure}[!htb]
\includegraphics[scale=0.60]{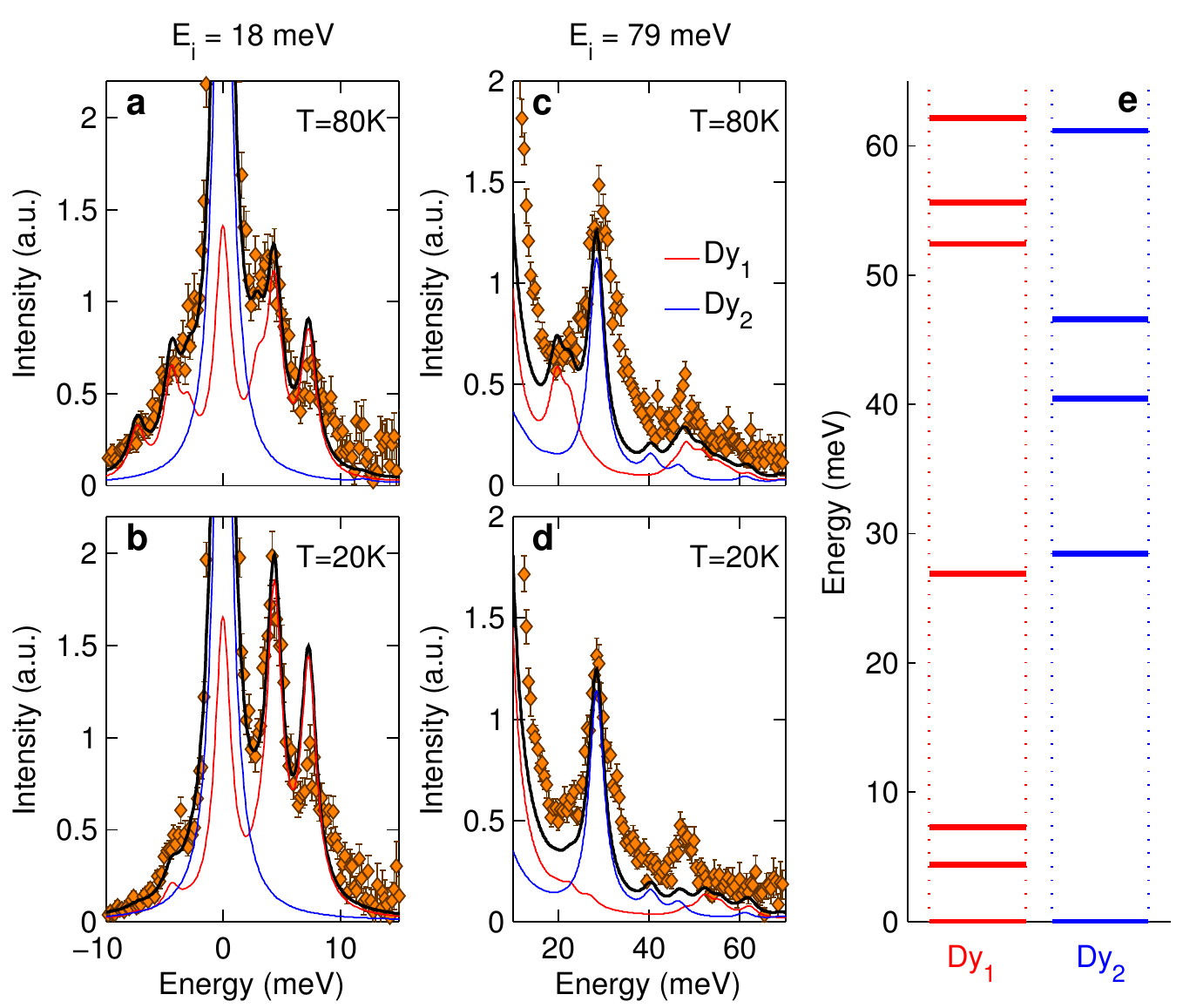}
\caption{(Color online) (a-d) Inelastic neutron spectra of \sdo\ for $E_i=18$~meV and $E_i=79$~meV at $T=20$~K and 80~K. The solid lines show the calculated spectra for both sites from the point charge calculation (see text) and (e) the corresponding energy levels scheme for both inequivalent Dy sites. The levels are doublet states.}
\label{HRPT/CEF}
\end{figure}

The results of the structural determination at $T=0.1$~K are presented in Table \ref{tablehrpt}. This low-temperature structure was used to revisit the crystal field level scheme presented by Fennell \textit{et al.}\cite{Fennell2014} In \sdo, the levels degeneracy is lifted by the $C_s$ symmetry at the Dy sites, resulting in a set of Kramers doublets. The excitations are modelled using a point charge calculation with crystal-field scaling factors $S_\text{xtal}^{1}=0.55$ and $S_\text{xtal}^{2}=0.41$ for sites 1 and 2 respectively.\cite{Uldry2012} By its nature, this model is extremely sensitive to the atomic positions. Figure \ref{HRPT/CEF} shows that the new structure gives a more appropriate description of the crystal-field excitations previously observed by Fennell \textit{et al.}\cite{Fennell2014} In particular, it reproduces the excitation at an energy transfer $E=8$~meV (Fig.~\refsub[b]{HRPT/CEF}) which was not accounted for in the previous fit. However, it does not reproduce well the intensity of the excitation at $E=48$~meV (Fig.~\refsub[d]{HRPT/CEF}). To verify the accuracy of this fit, calculations were done for slightly modified structures: spectra were calculated by moving single oxygen atom position within the uncertainty of the refined structure. The calculated scattering intensity is only weakly affected by this change but the energy levels change by up to 0.5~meV for the crystal-field excitations below $E=10$~meV and up to 1.5~meV for excitations above $E=10$~meV.

\begin{figure}
\begin{minipage}{.23\textwidth}
  \begin{tabular}{c|cc}
  \hline       \hline
     ~~~~                 &   ~~\textcolor{red}{Dy$_1$} ~~       & ~~\textcolor{blue}{Dy$_2$}~~       \\
\hline
 $g_1$           &   8.0               & 19.7             \\
 $g_2$           &   2.8               & $<0.1$               \\
 $g_3=g_c$   &   8.4               & $<0.1$               \\
 $\theta$        &   1.9$^\circ$   & 79.3$^\circ$ \\
   \hline       \hline   
\end{tabular}
\end{minipage}
\begin{minipage}{.23\textwidth}
  \includegraphics[width=0.85\textwidth]{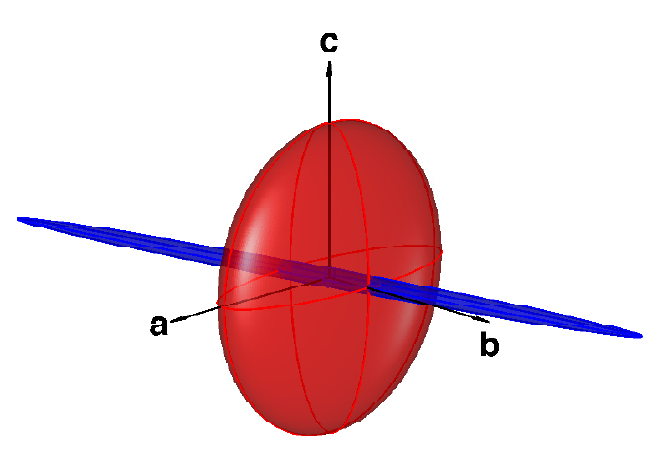}
\end{minipage}
\caption{(Color online) Left: $g$-factors along the principal axes calculated from the point charge calculations for both Dy sites. The principal axes of $g_1$ and $g_2$ are in the $ab$ plane and at $\theta$ and $\theta+90^\circ$ from the $a$-axis respectively. The principal axis of $g_3$ is along the $c$-axis. Right: Visual representation of the $g$-factors by ellipsoids for site 1 (red) and site 2 (blue).}
\label{table:gfactor}
\end{figure}

We calculated the gyromagnetic $\bf{g}$-tensor using the ground state doublet wavefunctions obtained from the experimentally determined crystal-field environment (see Appendix for details). The presence of a mirror symmetry in the $ab$ plane forces one of the principal axis of the gyromagnetic tensor to be along the $c$-axis and the two others are in the $ab$ plane. The calculated $g$-factors along the principal axes and their orientation are presented in Fig.~\ref{table:gfactor}. For site 2, the ground state doublet has a single easy-axis at 10.7$^\circ$ from the $b$-axis in the $ab$ plane, which is staggered in the lattice as shown on Fig.~\ref{fig:structure}. The values of the $g$-factor along the other principal axes are negligible making site 2 strongly Ising-like. For site 1, the $g$-factor along the $c$-axis is the largest but $g_1$ near the $a$-axis has a similar value, suggesting that site 1 is $XY$-like with an easy plane in the $ac$-plane. This $XY$-anisotropy is not protected by symmetry and contrasts with the usual Ising-anisotropy of Dy$^{3+}$ ions.\cite{Chilton2013} For both sites, there is no obvious correspondence between the suggested anisotropy and the oxygen atom positions surrounding the magnetic ions, an observation that also applies to SrEr$_2$O$_4$. \cite{Malkin2015}



We investigated the sensitivity of the $\bf{g}$-tensor to small variations of the oxygen atom positions within their experimental uncertainty. For site 2, the effects on the $g$-factors and the orientation of the principal axes were negligible. However, for site 1, the $g$-factors were strongly affected: $g_1$ and $g_c$ varied from $\sim$5 to 12 and the angle $\theta$ in the $ab$ plane took values up to 25$^\circ$, showing thus a high sensitivity to the crystal-field environment. This may lead to strong magneto-elastic effects. It also prevents a definitive description of the anisotropy on site 1: the moments could be $XY$-like or have an easy-axis along either the $a$ or $c$-axis. This result suggest that $XY$-anisotropy in Dy$^{3+}$ ions can generally arise 
at the crossover between two Ising-anisotropies with different directions.



\subsection{Zero field short range order}
\label{sec:ZF}

\begin{figure}[!htb]
\includegraphics{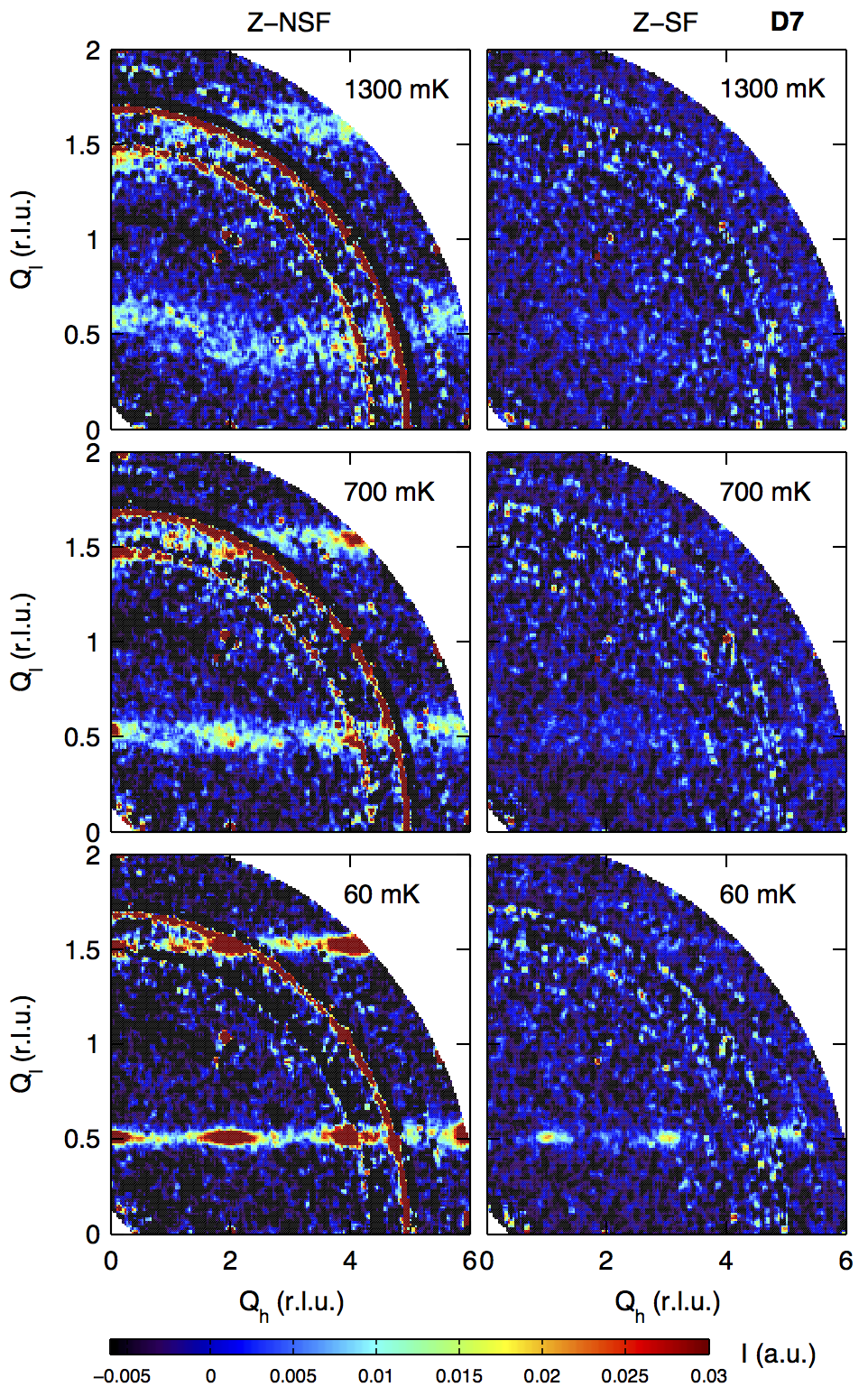}
\caption{(Color online) Diffuse NSF and SF scattering in the $(h0l)$ reciprocal plane at various temperature where the polarization Z is parallel to the $b$-axis. The powder rings in the NSF scattering originate from an incomplete background subtraction.}
\label{D7/D7ac}
\end{figure}

\begin{figure}[!htb]
\includegraphics{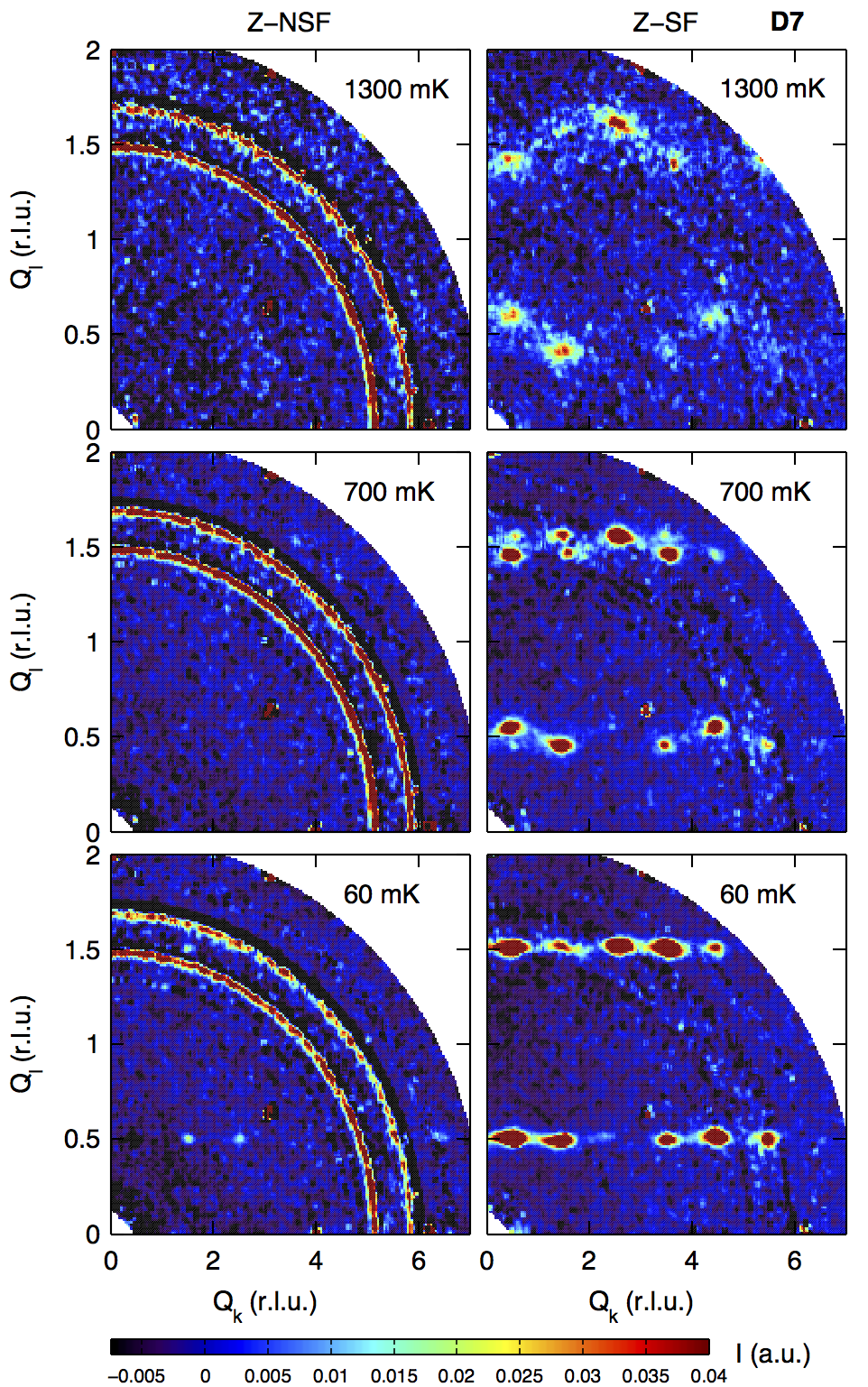}
\caption{(Color online) Diffuse NSF and SF scattering in the $(0kl)$ reciprocal plane at various temperature where the polarization Z is parallel to the $a$-axis. As the temperature increases the scattering moves to incommensurate positions. The powder rings in the NSF scattering originate from an incomplete background subtraction. The temperature-independent strong peaks appearing at incommensurate positions are nuclear Bragg peak of Si from the sample mount.}
\label{D7/D7bc}
\end{figure}

While there is no transition to long range magnetic order in \sdo , short range spin correlations gradually emerge with decreasing temperature. These short range spin correlations are clearly visible for two different sample orientations in the magnetic diffuse scattering measured on the D7 diffractometer with the polarization analysis. Neutrons were polarized along $Z$ perpendicular to the scattering plane and analyzed to separate the non-spin-flip (NSF) and spin-flip (SF) neutron scattering. The NSF scattering is sensitive to nuclear scattering and scattering from magnetic moments parallel to the neutron polarization. The SF scattering is sensitive to magnetic moments perpendicular to the neutron polarization. Figures \ref{D7/D7ac} and \ref{D7/D7bc} show the magnetic diffuse scattering in the $(h0l)$ and $(0kl)$ reciprocal plane respectively for both polarization channels at different temperatures. The data has been symmetrized into a single quadrant to improve statistics. 

At $T=60$~mK, lines of diffuse scattering are observed at half-integer values of $l$ in the $(h0l)$ plane in NSF scattering (Fig.~\ref{D7/D7ac}) and in the $(0kl)$ plane in SF scattering (Fig.~\ref{D7/D7bc}). These lines represent two-dimensional (2D) diffuse scattering planes close to $\textbf{Q}=(h,k,n+\frac{1}{2})$, where $n$ is an integer. In real space, this indicates the presence of 1D correlations along the $c$-axis, which is the direction of the zig-zag chains. With increasing temperature, the diffuse scattering is modulated around $l=\nobreak n+\frac{1}{2}$ as a function of $h$ and $k$. As a consequence, the diffuse scattering forms distorted 2D planes at $T=1.3$~K.

The polarization analysis of the diffuse scattering indicates that the moments giving rise to this scattering predominantly lie along the $b$-axis. For the crystal orientation with the $b$-axis vertical, the strong NSF scattering in the $(h0l)$ reciprocal plane near even $h$ values must originate from moments along the $b$-axis. Weaker intensity of the SF scattering appears near odd $h$ values at the lowest temperature, indicating a smaller moment component in the $ac$ plane. For the crystal orientation with the $a$-axis vertical, strong SF intensity is observed in the $(0kl)$ reciprocal plane near half-integer values of $k$, indicating moments in the $bc$ plane. A small moment component along the $a$-axis also emerges at $T=60$~mK, as seen from the NSF scattering near $\textbf{Q}=(0,1.5,0.5)$ and $(0,2.5,0.5)$ shown in Fig.~\ref{D7/D7bc}. 

\begin{figure}[!htb]
\includegraphics[scale=1.1]{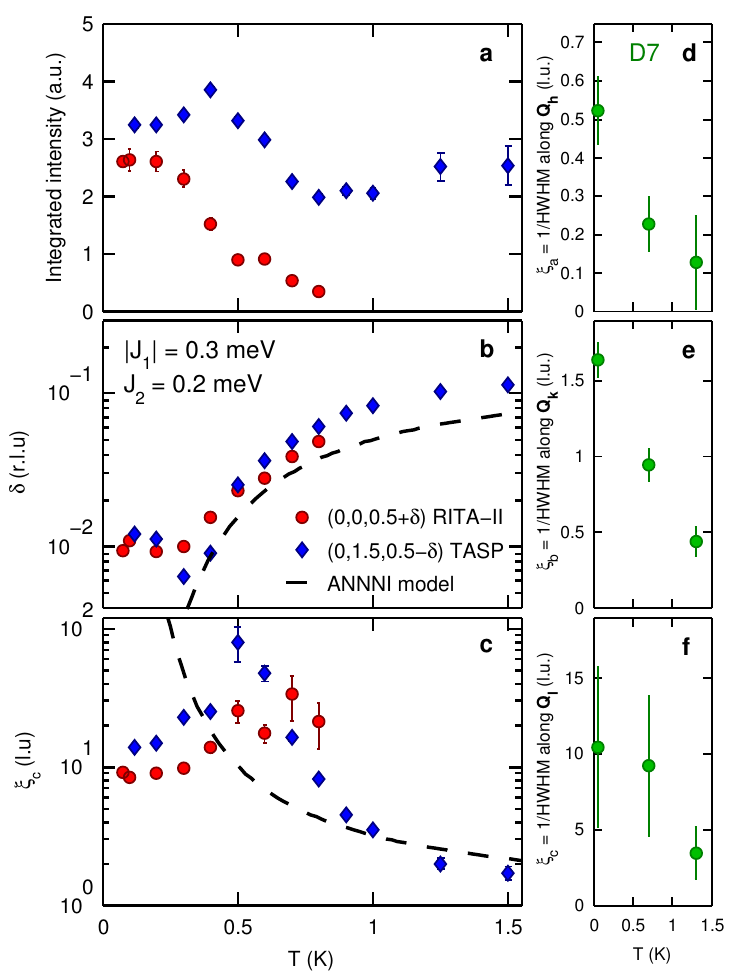}
\caption{(Color online) Temperature dependence of the diffuse scattering at $\textbf{Q}=(0,0,0.5+\delta)$ and $(0,1.5,0.5-\delta)$, showing (a) the integrated intensity, (b) the $\delta$ parameter of the incommensurability and (c) the peak width along $l$ presented as the correlation length $\xi$ along the $c$-axis. The analytical functions for $\delta$ and $\xi$ from the 1D ANNNI model with $|J_1|=0.3$~meV and $J_2=0.2$~meV are shown as dashed lines. (d-f)~Correlation length represented as the inverse half width at half maximum (HWHM) of the diffuse scattering along $h$, $k$ and $l$ measured on the D7 diffractometer.}
\label{RITA/TASP_RITA_dataComparison}
\end{figure}

A detailed temperature dependence of the diffuse scattering for $T<1.5$~K was measured near $\textbf{Q}=(0,0,0.5 + \delta)$ using the RITA-II spectrometer and near $\textbf{Q}=(0,1.5,0.5 - \delta)$ using the TASP spectrometer (Fig.~\ref{RITA/TASP_RITA_dataComparison}). The integrated intensity, the correlation length and the wave-vector number $\delta$ were determined by fitting a Lorentzian function to the diffuse scattering measured along $l$. The correlation length was corrected for the wave-vector resolution of the instrument and is defined as $\xi_c = 2/\sqrt{\Gamma_\text{exp}^2-\Gamma_\text{res}^2}$ where $\Gamma_\text{exp}$ is the full-width at half maximum (FWHM) of the measured scattering and $\Gamma_\text{res}$ is the FWHM expected from the instrumental resolution. The instrumental resolution was calculated in the Popovici approximation. \cite{Popovici1975}

From $T=1.5$~K down to 0.5~K, there is an increase of the correlation length along the $c$-axis. Simultaneously with the increasing correlation length, there is a reduction of the incommensurate wave-vector number $\delta$. This shows that the magnetic correlations converge towards long range commensurate order with decreasing temperature. However, the incommensuration becomes less temperature dependent below $T=0.5$~K and remains around $\delta \approx 0.01$. The correlation length surprisingly decreases with decreasing temperature below $T=0.5$~K. This is clearly visible for the correlation length determined near $\textbf{Q}=(0,1.5,0.5)$ that corresponds to the putative ordering wave-vector. This identifies $T=0.5$~K as a temperature where the magnetic properties change qualitatively. 

In \seo\ and \sho, the value $\delta$ of the diffuse scattering also does not reach zero at the lowest temperatures.\cite{Hayes2011,Wen2015} For \sho\ in particular, $\delta$ stabilizes at 0.001 below $T=0.52$~K, but no reduction of the correlation length is observed.\cite{Wen2015} This behaviour is explained by the interaction between the short range correlations on site 2 and the long range order on site 1. However, such a scenario can not explain the magnetic phenomena reported here in \sdo\ because no magnetic long range order is present at zero field at any temperature. 

The half-width at half maximum (HWHM) was also determined for the temperatures measured with the D7 diffractometer. The average HWHMs along $k$ and $l$ have been obtained from the SF scattering in the $(0kl)$ reciprocal plane from all the peaks at $l=0.5 \pm \delta$. The average HWHMs along $h$ have been obtained from the NSF scattering in the $(h0l)$ reciprocal plane from all the peaks at $l=0.5 \pm \delta$. These results are presented on Fig.~\refsub[d-f]{RITA/TASP_RITA_dataComparison} as the correlation length $\xi \approx 1/\text{HWHM}$. Here we assumed that the instrumental resolution is much smaller than the experimental widths. For these three temperatures, the correlation lengths determined from the D7 data agree with those obtained from TASP. New information is obtained from the temperature dependence of the correlation length along the $a$ and $b$ axes. Along the $b$-axis, the correlation length is short but finite at $T=1.3$~K and increases continuously down to 60~mK. Along the $a$-axis, the correlation length is negligible at $T=1.3$~K and only becomes significant between 700~mK and 60~mK. This means that although \sdo\ is in first approximation a 1D magnet, it features 2D correlations at $T=1.3$~K and a crossover to 3D correlations between $T=700$~mK and 60~mK.

We used the SF scattering in the $(0kl)$ reciprocal plane to determine the nature of the correlations with a propagation vector $\textbf{k}_{1/2}=(0,\frac{1}{2},\frac{1}{2})$. The integrated intensities of the short range peaks at $(0,\frac{m}{2},\frac{n}{2})$ for $T=60$~mK were obtained by summing up the counts over $\Delta l =0.4$ and fitting the $k$-dependence of the resulting intensity with Lorentzian functions. The integrated intensities  were corrected for absorption, which was estimated through finite element analysis based on the sample geometry, and then used to determine the nature of the short range correlations. Since these intensities arise from SF scattering of neutrons polarized along the $a$-axis, they are only sensitive to magnetic moments in the $bc$ plane. The NSF scattering in the $(0kl)$ reciprocal plane is necessary to describe the components along the $a$-axis.

\begin{table}
\begin{tabular}{c|cc}
  \hline      
     ~~~~~~~~~~~                 &  ~~~~~~~~~~~~~~$\Gamma_1$ ~~~~~~~~~~~~~~       & ~~~~~~~~~~~~~~$\Gamma_2$~~~~~~~~~~~~~~       \\
\hline
 $x,y,z$ &              \begin{tabular}{@{}cc@{}}  $(+A_x,+A_y,+A_z)$  \\ $(-B_x,-B_y,+B_z)$ \end{tabular} &
 			    \begin{tabular}{@{}cc@{}}  $(+A_x,+A_y,+A_z)$  \\ $(-B_x,-B_y,+B_z)$ \end{tabular} \\
\hline                                                                                                       
 $x+\frac{1}{2},-y+\frac{1}{2},-z+\frac{1}{2}$ & 
			 \begin{tabular}{@{}cc@{}}  $(-A_x,+A_y,+A_z)$  \\ $(+B_x,-B_y,+B_z)$ \end{tabular} &
 			 \begin{tabular}{@{}cc@{}}  $(+A_x,-A_y,-A_z)$  \\ $(-B_x,+B_y,-B_z)$ \end{tabular} \\
\hline                                                                                                       
  $-x+1,-y+1,z+\frac{1}{2}$ & 
			  \begin{tabular}{@{}cc@{}}  $(+A_x,+A_y,+A_z)$  \\ $(+B_x,+B_y,-B_z)$ \end{tabular} &
 		          \begin{tabular}{@{}cc@{}}  $(+A_x,+A_y,+A_z)$  \\ $(+B_x,+B_y,-B_z)$ \end{tabular} \\
\hline                                                                                                       									
   $-x+\frac{1}{2},y+\frac{1}{2},-z+1$ & 
		             \begin{tabular}{@{}cc@{}}  $(-A_x,+A_y,+A_z)$  \\ $(-B_x,+B_y,-B_z)$ \end{tabular} &
 			    \begin{tabular}{@{}cc@{}}  $(+A_x,-A_y,-A_z)$  \\ $(+B_x,-B_y,+B_z)$ \end{tabular} \\
    \hline      
    \end{tabular}
\caption{The six basis vectors represented as $A_{x,y,z}$ and $B_{x,y,z}$ are listed for the two 2D irreducible representations with the propagation vector $\textbf{k}_{1/2}=(0,\frac{1}{2},\frac{1}{2})$ at Wyckoff position $4c$ in space group \it{Pnam}.}
\label{table:irreps}
\end{table}

\begin{figure}[!htb]
\includegraphics[scale=0.9]{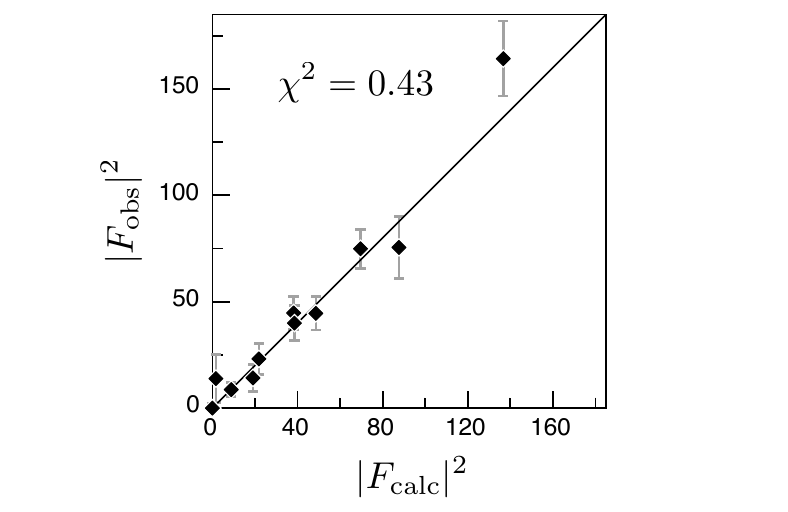}
\vspace{-0.4cm}
\caption{Result of the magnetic structure refinement at $T=60~$mK, represented as the observed squared structure factors $|F_\text{obs}|^2$ obtained from the diffuse scattering (see text) versus the calculated squared structure factors $|F_\text{calc}|^2$.}
\label{D7/refinement}
\end{figure}

To determine the symmetry of the short range magnetic correlations, the representation analysis for $\textbf{k}_{1/2}$ has been performed using \textsc{BasIreps} \cite{Fullprof} and SARAh.\cite{Wills2000} The little group of $\textbf{k}_{1/2}$ contains two 2D irreducible representations, both having six basis vectors represented as $A_{x,y,z}$ and $B_{x,y,z}$, where the indices refer to the crystallographic axes (Table \ref{table:irreps}). By restricting the refinement to the SF scattering, only moments in the $bc$ plane are relevant. The number of refined parameters was further reduced by assuming that the moment on site 1 points along $c$ and the one on site 2 points along $b$. This assumption is supported by the anisotropy of the $\bf{g}$-tensor of both sites presented in section \ref{sec:structure}. Therefore, only four parameters are adjustable: $A_z$, $B_z$ on site 1, which we denote as $A_z^1$ and $B_z^1$,  and $A_y$, $B_y$ on site 2, which we denote as $A_y^2$ and $B_y^2$. Assuming that all moments on site 1, and all moments on site 2, have the same size, the model is restrained to one free parameter per inequivalent site ($A_z^1$ or $B_z^1$ must be zero for site 1, and $A_y^2$ or $B_y^2$ must be zero for site 2). Based on these assumptions, a good agreement with the data is obtained from the coexistence of two domains generated from different basis vector of the $\Gamma_2$ representation. One domain is described by ($A_z^1,A_y^2$) and the other one by ($B_z^1,B_y^2$). We further assumed that the modulus of $A_z^1$ and $B_z^1$ are the same for both domains, as well as for $A_y^2$ and $B_y^2$. Assuming an equal population of domains, this magnetic structure refinement gives $\chi^2=0.43$. The refinement was performed using \textsc{FullProf} \cite{Fullprof} and the result is presented in Fig.~\ref{D7/refinement}. The moment size on site 2 along the $b$-axis is about twice the one on site 1 along the $c$-axis, in good agreement with the expectation from the $g$-factors. The absolute moment sizes were not obtained because the nuclear peaks intensity could not be extracted accurately. 

The NSF scattering observed in the $(0kl)$ reciprocal plane provides information about the magnetic correlations polarized along the $a$-axis. The presence of weak intensity near $\textbf{Q}=(0,1.5,0.5)$ and $(0,2.5,0.5)$ in NSF scattering provides evidence for such correlations. For the $\Gamma_2$ representation, this scattering is attributed unambiguously to a moment along the $a$-axis on site 2, which is consistent with the easy-axis of the calculated $\bf{g}$-tensor. A moment along the $a$-axis on site 1 would lead to NSF scattering at $\textbf{Q}=(0,0.5,0.5)$, which is experimentally not observed. This is evidence that there is no ordered moment along the $a$-axis on site 1, although the calculated $\bf{g}$-tensor would allow this. This indicates that either the $XY$ anisotropy on site 1 at low temperatures is weak, or that the ordering of this spin component is frustrated by the interactions.

\begin{figure}[!htb]
\includegraphics[scale=1.1]{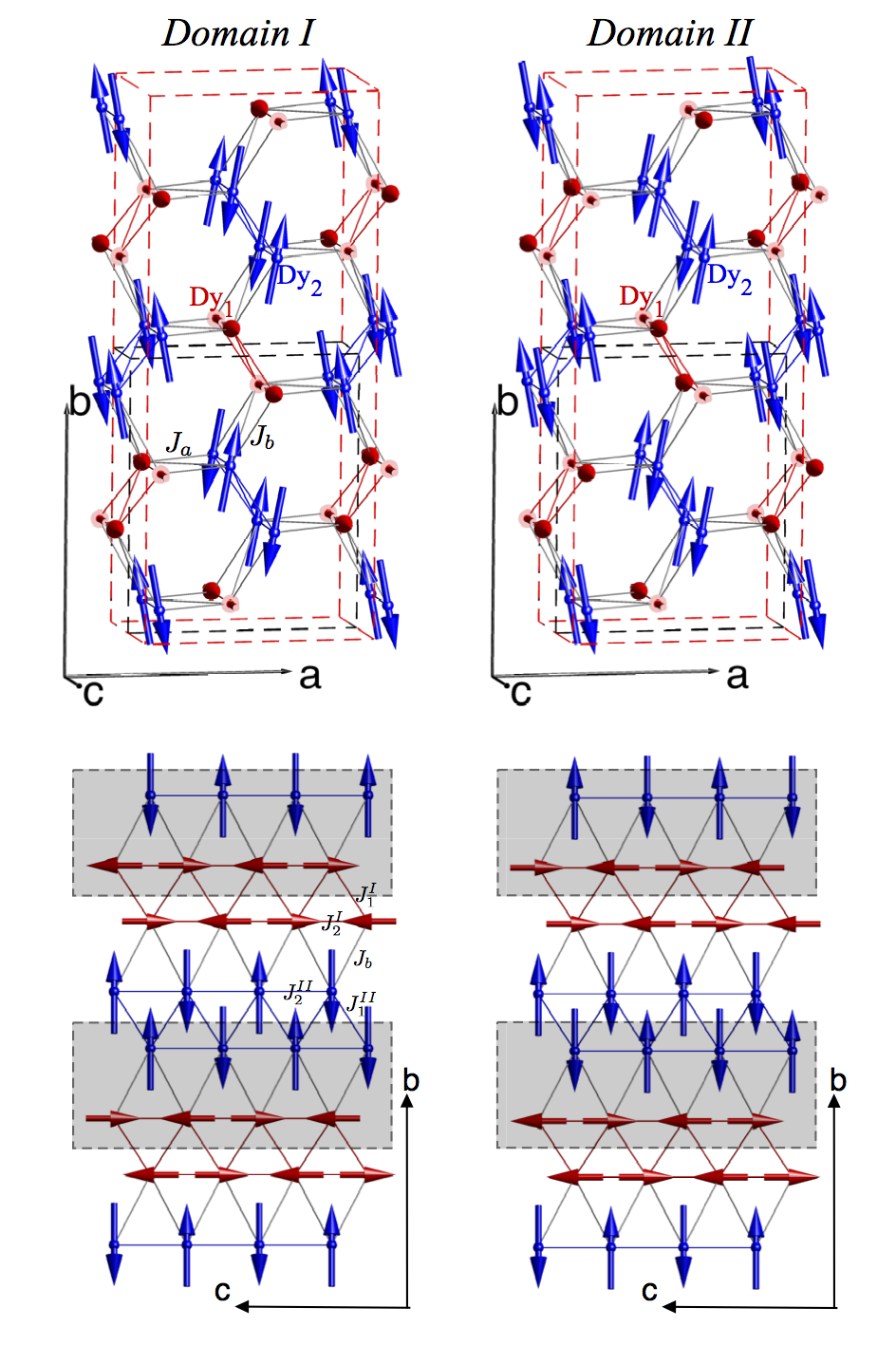}
\caption{(Color online) Magnetic structure of the coexisting domains in the zero-field short range order of \sdo\ with $\textbf{k}_{1/2}$. The chemical unit cell is outlined in gray lines and the magnetic unit cell in red lines. The shaded areas shown in the $bc$ plane enlighten the difference between both domains. Moments on site 1 point along the $c$-axis and moments on site 2 point nearly along the $b$-axis in the $ab$ plane. }
\label{D7/structure}
\end{figure}

The short range correlations at $T=60$~mK are well described by the coexistence of the two resulting arrangements of magnetic moments shown in Fig.~\ref{D7/structure}. These spin correlations decay with a correlation length of about 20 sites along the $c$-axis, 3 sites along the $b$-axis and 1 site along the $a$-axis. The two arrangements are very similar, showing a double N\'eel state on both chains, and could be considered as different domains if they were long ranged ordered. The difference between the two arrangements appears in the shaded area illustrated in the $bc$ plane in Fig.~\ref{D7/structure}: all the moments in the non-shaded area point in same direction for both arrangements while the shaded ones point in opposite direction in one arrangement relative to the other. Both arrangements must be present in equal amount to account for the observed scattering. This reflects a decoupling of the moments in the shaded area relative to the ones in the non-shaded area, indicating two independent sublattices.

The diffuse scattering in the $(h0l)$ reciprocal plane was not used to determine the nature of the spin correlations. This is because the propagation vector $\textbf{k}_{1/2}$ lies outside this plane. Nevertheless, the observed scattering in the $(h0l)$ reciprocal plane also arises from the magnetic correlations described by $\textbf{k}_{1/2}$ because the correlation length along the $b$-axis is short and leads to peak broadening in reciprocal space along $k$. The magnetic scattering near $\textbf{Q}=(h,0,l)$ originates from $\textbf{Q}=(h,-0.5,l)$ and $(h,0.5,l)$. At these positions, the model predicts strong NSF scattering for even $h$ values and strong SF scattering at odd $h$ values, in agreement with the observations in the $(h0l)$ reciprocal plane. 

\begin{figure}[!htb]
\includegraphics{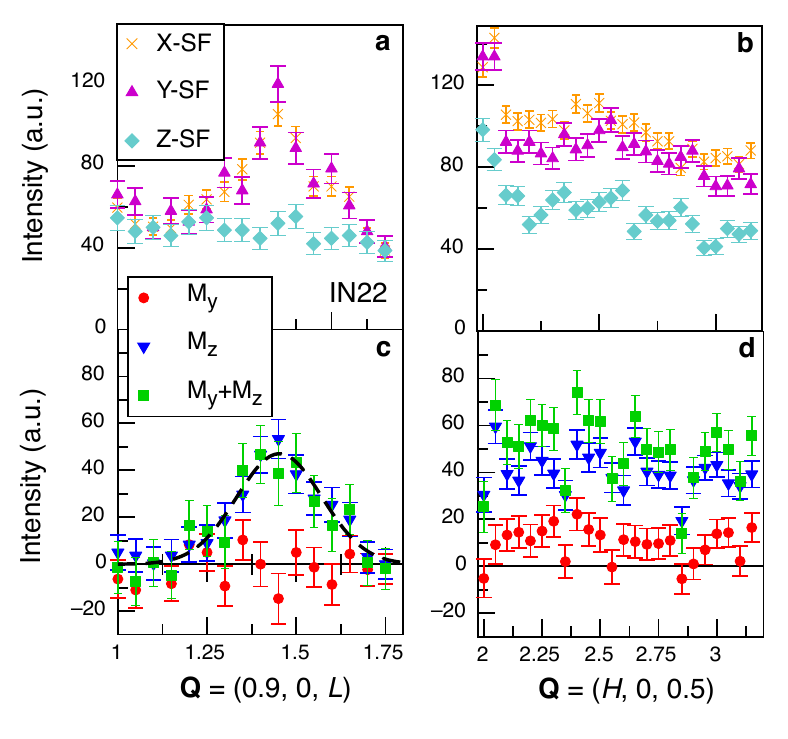}
\vspace{-0.5cm}
\caption{(Color online) \textbf{Q}-scan of the magnetic diffuse scattering along $h$ and $l$ at $T=1.4$~K on the IN22 spectrometer (a-b) measured by spin-flip scattering polarized along $X$, $Y$ and $Z$ and (c-d) the extracted magnetic vector components $M_y$, $M_z$ and $M_y+M_z$.}
\label{IN22_MagScattPol}
\end{figure}

Detailed polarized measurements of the diffuse scattering have also been measured on the IN22 spectrometer at $T=1.4$~K (Fig.~\ref{IN22_MagScattPol}) and show that already at this temperature both inequivalent sites contribute to the magnetic correlations. The magnetic diffuse SF scattering was measured for neutrons polarized along $X$, $Y$ and $Z$, where $X$ is parallel to $\textbf{Q}$, $Y$ is perpendicular to $X$ in the scattering plane and $Z$ is perpendicular to the scattering plane. These different cross-sections were combined to remove the nuclear spin-incoherent background and extract the components of the magnetic scattering vector \cite{Regnault2006}: $M_y+M_z=2\sigma_x^\text{SF}-\sigma_y^\text{SF}-\sigma_z^\text{SF}$, $M_y=\sigma_x^\text{SF}-\sigma_y^\text{SF}$ and $M_z=\sigma_x^\text{SF}-\sigma_z^\text{SF}$. Since the sample was oriented in the $ac$ scattering plane, $M_z$ is the magnetic component along the $b$-axis while $M_y$ measures contribution of components along the $a$ and $c$ axes. The sensitivity of $M_y$ to these two contributions depends on the direction of $\textbf{Q}$.

Near $\textbf{Q}=(0.9,0,1.45)$, magnetic scattering is only observed for $M_z$, indicating a strong component along the $b$-axis and implying dominant correlations on site 2 (Fig.~\refsub[c]{IN22_MagScattPol}). Along $\textbf{Q}=(h,0,0.5)$, the strongest scattering is observed for the $M_z$ channel but there is a also non-zero contribution in $M_y$, which is mostly sensitive to moments along the $c$-axis that are related to site 1. This shows that the diffuse scattering arises from magnetic correlations on both site 1 and 2. The presence of the same correlations on both inequivalent sites distinguishes \sdo\ from the analogous compounds \sho\ and \seo\ where both inequivalent sites have different orderings and temperature dependences.\cite{Young2013,Wen2015,Hayes2011}

\subsection{AC magnetic susceptibility}
\label{sec:AC}
\begin{figure}[!htb]
\includegraphics[scale=1.1]{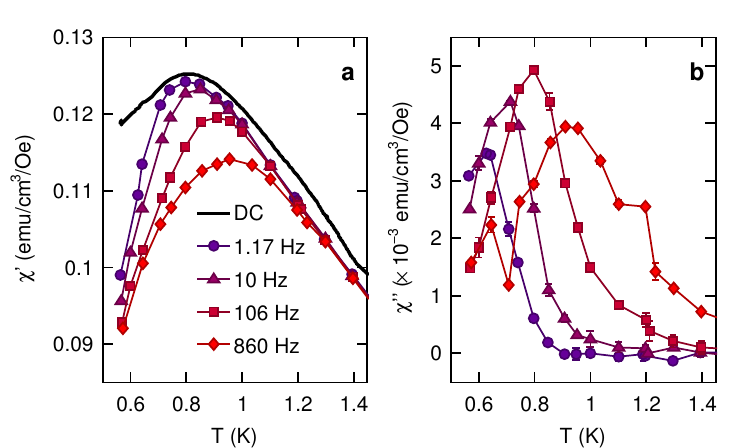}
\label{XT}
\caption{(Color online) AC susceptibility measurements showing the temperature dependence of (a) the real part $\chi'$ and (b) the imaginary part $\chi''$ at different excitation frequencies of $H=1$~Oe applied along the $b$-axis. In the limit of low frequency, $\chi'$ converges to the DC susceptibility.}
\end{figure}

The real part of the AC susceptibility $\chi'$ and the imaginary part $\chi''$ are presented as function of temperature in Fig.~\ref{XT} and as a function of frequency in Fig.~\refsub[a-b]{ACchi}. There is a maximum in $\chi'$ at $T=0.8$~K that shifts to higher temperatures with increasing frequency. This maximum as function of temperature is also visible in the DC susceptibility shown in Fig.~\refsub[a]{XT} for comparison. A frequency-dependent maximum in the susceptibility is often observed in spin glass systems where the fluctuation rate is slow. Spin freezing in frustrated magnets like the spin ices also show similar dependencies on the frequency.\cite{Lhotel2012} 

We also observed broad peaks in the frequency dependence of $\chi''$, indicating a characteristic relaxation time longer than 1~ms below $T\approx1~$K. The relaxation time given by the $\chi''$ peak maximum follows an Arrhenius law $\tau=\tau_0 \exp(E_a/k_\text{B}T)$ as shown on Fig.~\refsub[d]{ACchi}. This indicates thermally activated processes with an activation energy $E_a \approx 9.4$~K and a characteristic time $\tau_0 \approx 7.1 \times 10^{-8}$~s. The lowest crystal-field level occurs at 4~meV~($\sim45$~K) and the energy scale of the magnetic interactions is about 3~K (see section \ref{sec:ANNNI}). This suggests that the activation energy results from multiple spin-flip processes governed by magnetic interactions.

\begin{figure}[!htb]
\includegraphics[scale=1.1]{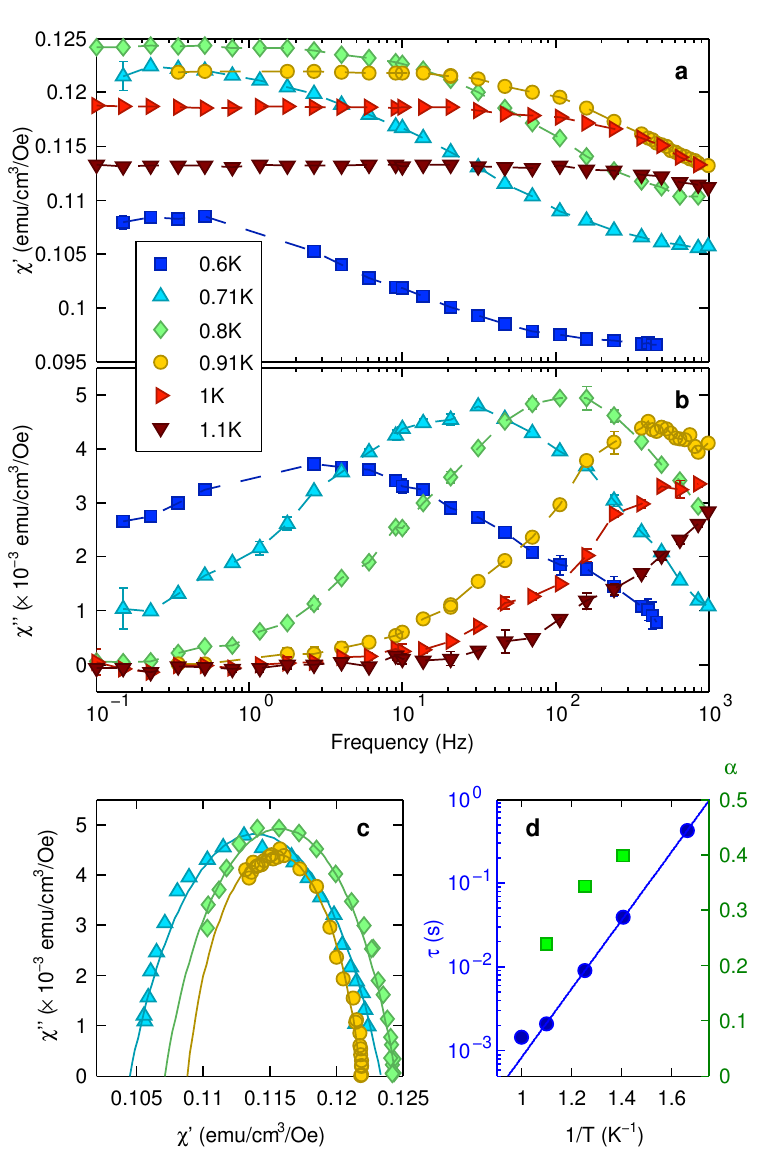}
\label{ACchi}
\caption{(Color online) Frequency dependence of (a) the real part of the AC susceptibility $\chi'$ and (b) the imaginary part $\chi''$ from $T=0.6$ to 1.1~K with an excitation amplitude of 1~Oe. (c) $\chi''$ as function of $\chi'$ for $T=0.71$, 0.8 and 0.91~K fitted using the Cole-Cole formalism (Eq.~\ref{colecole}). (d) Temperature dependence of the relaxation peaks maximum fitted with the Arrhenius equation (blue circle) and temperature dependence of the fitted parameter $\alpha$ of the Cole-Cole function (green square).}
\end{figure}

The peak observed in the frequency dependence of $\chi''$ is broader at $T=0.6$~K compared to $T=0.8$~K, indicating a larger distribution of relaxation times below $T=0.5$~K. This can be quantified by tracing $\chi''$ as a function of $\chi'$, which should be semi-circular for a single relaxation time (usually called Debye relaxation) but is flattened in the presence of a distribution of relaxation times. This is described in the Cole-Cole formalism by
\begin{equation}
\begin{aligned}
\chi'' & (\chi')= -\frac{\chi_0-\chi_s}{2\tan \left[ (1-\alpha) \frac{\pi}{2} \right]} \\
& + \sqrt{(\chi'-\chi_s)(\chi_0-\chi')+\frac{(\chi_0-\chi_s)^2}{4 \tan^2 \left[ (1-\alpha) \frac{\pi}{2} \right]}}
\end{aligned}
\label{colecole}
\end{equation}
where $\chi_0$ is the susceptibility for very low frequencies, $\chi_s$ for very high ones and $\alpha$ is a parameter related to the distribution of relaxation times.\cite{Hagiwara1998} The Debye relaxation with a single relaxation time is characterized by $\alpha=0$ while non-zero values of $\alpha$ represent a distribution of relaxation times which is infinitely broad at $\alpha=1$. Plots of $\chi''(\chi')$ at $T=0.71$~K, 0.80~K and 0.91~K are shown in Fig.~\refsub[c]{ACchi} and the fitted $\alpha$ values in Fig.~\refsub[d]{ACchi}. For these temperatures, a single relaxation time is not sufficient to describe the data, confirming the existence of a distribution of relaxation times which broadens with decreasing temperature. 

\section{Discussion}
\subsection{ANNNI model}
\label{sec:ANNNI}
It has been previously pointed out that \sdo\ and other members of the family can be described by two inequivalent zig-zag chains with Ising-like spins.\cite{Fennell2014} The Hamiltonian of a single Ising zig-zag chain is that of the 1D axial next-nearest neighbour Ising (ANNNI) model, which is given as
\begin{equation}
{H} = \sum_i  J_1 {\hat{S}_i^z} {\hat{S}_{i+1}^z} + J_2 {\hat{S}_i^z} {\hat{S}_{i+2}^z}. 
\end{equation}
Here $J_1$ is the nearest-neighbour interaction and $J_2$ is the next-nearest-neighbour interaction. This model is exactly solvable in zero field.\cite{Selke1988} The correlations of the \uudd\ type observed by neutron scattering in \sdo\ correspond to the double N\'eel ground state, which is predicted in the 1D ANNNI model for ${J_2}/{|J_1|}>0.5$. Furthermore, a magnetization plateau at $\frac{1}{3}M_s$ is stabilized only for antiferromagnetic $J_1>0$ in this model. The presence of such plateau for a field applied along the $b$-axis at $T=0.5~K$ indicate an antiferromagnetic $J_1$ interaction on site~2, where moments lie close to the $b$-axis. The interactions for site 2 can be estimated from the critical fields $H_p=0.16$~T at the plateau onset and $H_s=2.03$~T at the saturation.\cite{J.Hayes2012} The interactions are given by $J_1=M(H_s-H_p)/3=0.35$~meV, $J_2=M(H_s+2H_p)/6=0.22$~meV and $J_2/J_1=0.63$, assuming the moment $M$ to be the $b$-component of site 2 easy-axis. 

The spin-spin correlation function $G(r)$ was calculated analytically for the 1D ANNNI model by Stephenson \cite{Stephenson1970} and for ${J_2}/{|J_1|}>0.5$, it can be expressed in the simple form:
\begin{equation}
G(r)=A e^{-\frac{r}{\xi}} \cos(2\pi qr+\phi)
\label{correlationfunction}
\end{equation}
where $r$ is the spin site, $A$ is a scaling factor, $\xi$ is the correlation length, $q$ is the wavenumber and $\phi$ is a phase factor. All these parameters have complex dependence on $J_1$, $J_2$ and the temperature. Neutron scattering probes directly the wavenumber $q$ and the correlation length $\xi$. Therefore, the experimental values of these parameters can be compared with the analytical expressions for the 1D ANNNI model. The experimental temperature dependences of the incommensurability and the correlation length are compared with the expectations of the 1D ANNNI model for $|J_1|=0.3 $~meV and $J_2=0.2$~meV in Fig.~\refsub[b-c]{RITA/TASP_RITA_dataComparison}. The interactions determined from the critical fields of the magnetization plateau were used as starting parameters, and only a qualitative agreement of both the incommensurability and the correlation length with the model can be achieved by adjusting the strength of the interactions. The sign of $J_1$ does not affect the temperature dependence of $q$ and $\xi$ for $J_2/|J_1|>0.5$ as it is the case here. It is assumed that $S_i=\pm1$, leading to $J_{ij}|S_iS_{j}|=J_{ij}$. Below $T=0.5$~K, both parameters diverge significantly from the model, which we attribute to a dimensionality crossover, as will be discussed in the next section. 

\begin{figure}[!htb]
\includegraphics[scale=0.55]{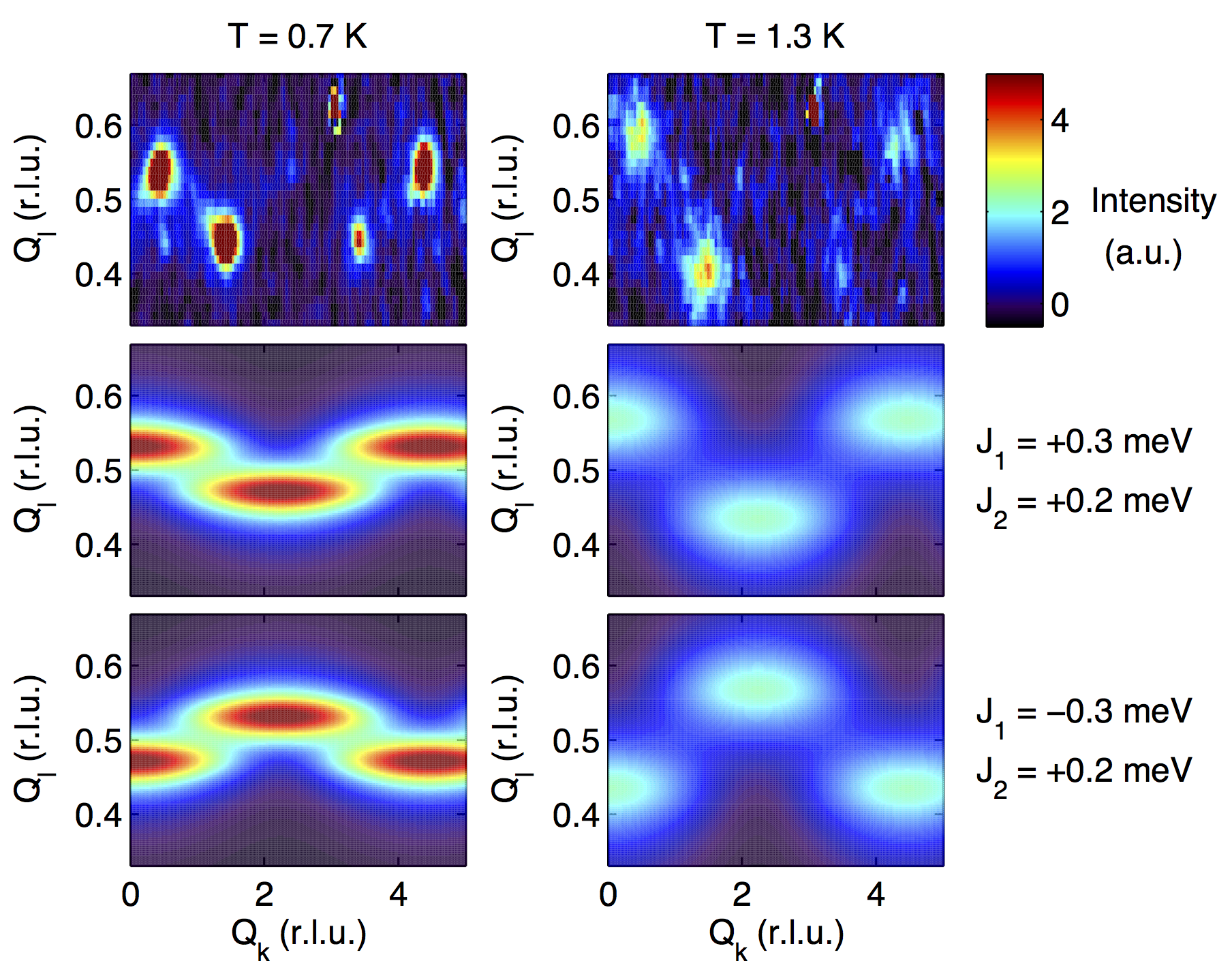}
\vspace{-0.3cm}
\caption{(Color online) Comparison between the experimental diffuse scattering in the $(0kl)$ reciprocal plane and the Fourier transform of the 1D ANNNI model correlation function (Eq.~\ref{correlationfunction}) on the zig-zag chain structure for $J_1=\pm0.3$~meV and $J_2=+0.2$~meV.}
\label{D7/calculation}
\end{figure}

The position of the diffuse scattering maximum in $l$ ($+\delta$ or $-\delta$) is dependent on $h$ and $k$, suggesting that it arises from the presence of 3D correlations. However, this shift in reciprocal space is attributed to the geometry of the zig-zag chains in the presence of 1D correlations only. This observation has previously also been made in \sho\ by Wen \textit{et al.}\cite{Wen2015} The Fourier transform of the correlation function (Eq. \ref{correlationfunction}) on the zig-zag chain has been calculated for \{$J_1=+0.3$, $J_2=+0.2$~meV\} and \{$J_1=-0.3$, $J_2=+0.2$~meV\} and compared to the experimental results (Fig.~\ref{D7/calculation}). The alternation of the scattering maximum position above and below $l=0.5$ as function of $k$ is well taken into account when $J_1$ is antiferromagnetic, while it is inverted for a ferromagnetic $J_1$. The same applies as a function of $h$ (not shown). This comparison therefore confirms that $J_1$ must be antiferromagnetic, as expected from the presence of the magnetization plateau. Even though this alternation of the maximum position around $l=0.5$ does not indicate 3D correlations, the amplitude modulation along the $h$ and $k$ directions are signatures of interchain interactions which are discussed in the section \ref{sec:crossover}.

The diffuse scattering shown in Fig.~\ref{D7/calculation} is relatively well described by a single type of correlations originating from a zig-zag chain with effective interactions $J_1=+0.3$~meV and $J_2=+0.2$~meV. We can therefore speculate that both inequivalent sites have similar interactions. The neutron polarization analysis also confirms that both sites contribute to the observed correlations up to $T=1.4$~K. Similar but not identical interactions on both inequivalent chains could cause additional frustration in this system. The presence of a single type of correlations can alternatively be understood by strong interactions on one inequivalent chain and weak interactions on the other one. In that case, one chain would dominate the correlations and the other chain would be driven by interchain interactions. However, this would mean that the energy scale on one chain is much larger than on the other, which is inconsistent with the magnetization data showing similar saturation fields for the two sites.\cite{J.Hayes2012}  


%
Assuming that both inequivalent chains have similar interactions, the magnetization plateau at $\frac{1}{3}M_s$ that is predicted for $J_2/|J_1|>0.5$ should be observed for both sites. It is stabilized for fields applied along the $b$-axis and can be related to site 2 from its easy-axis. However, there is no clear observation of a magnetization plateau for fields applied along the $c$-axis,\cite{J.Hayes2012} the direction of the ordered moment on site 1. This may arise from a weak $XY$ anisotropy on site 1 with an easier axis along the $c$-axis than along the $a$-axis, as suggested by magnetization measurements.\cite{J.Hayes2012} 

\subsection{Dimensionality crossover}
\label{sec:crossover}

In order to understand the system dimensionality, it is important to establish the possible interaction pathways, especially for the interchain interactions. The shortest bonds between zig-zag chains appear between the two inequivalent sites and are probably the dominant interchain interactions. A possible interaction mechanism is the isotropic exchange interaction which couples together the moment components along the same direction. However, since our results suggest that the effective spin direction on the inequivalent sites are orthogonal, the isotropic exchange is expected to be weak between the zig-zag chains. The interchain coupling could therefore originate from anisotropic interactions such as Dzyaloshinskii-Moriya, dipolar and/or multipolar interactions. Concerning the Dzyaloshinskii-Moriya interaction $\left(\textbf{D}_{ij}=\left( \textbf{S}_i \times \textbf{S}_j \right) \right)$, the symmetry allows only a $D^z_{ij}$ component for the nearest-neighbour interchain bonds, coupling moment components in the $ab$ plane. Since our results indicate that the moment on site 1 has no significative component in this plane, this interaction probably does not contribute to the interchain coupling. 

The presence of large moments suggest that the dipolar interactions will be relevant. Assuming the moments from the easy-axis of the $g$-factors ($\bf{M}\rm{_1=(0,0,4.2)\mu_\text{B}}$ and $\bf{M}\rm{_2=(1.8,9.7,0)\mu_\text{B}}$),  the strongest interchain dipolar interaction is $J_b=0.0376$~meV, coupling the zig-zag chains along the $b$-axis (Fig.~\ref{D7/structure}). All the other dipolar interchain interactions are at least twice weaker. The strongest coupling along the $a$-axis is $J_a=0.0086$~meV as shown in Fig.~\ref{D7/structure}. This suggest that stronger correlations are expected along the $b$-axis than the $a$-axis. This is indeed what is observed with a correlation length at least twice as large along the $b$-axis than along the $a$-axis (Fig.~\refsub[d-e]{RITA/TASP_RITA_dataComparison}). This qualitative agreement indicates that the dipolar interactions are likely the dominant interchain interactions. The energy of the $J_a$ and $J_b$ interactions is minimized by the arrangements of magnetic moments of the correlations described in section \ref{sec:ZF}, supporting the conclusion that the dipolar interactions play an important role. 

Since the correlation length is much longer along the chain ($c$-axis) than along the $a$ and $b$ axes at $T=1.3$~K, the system can be described as 1D zig-zag chains in first approximation. This explains why the temperature dependence of the incommensurability parameter $\delta$ and the correlation length $\xi_c$ is well described by the 1D ANNNI model above $T=0.7$~K. However, the presence of an intensity modulation along $k$ at $T=1.3$~K shows that the correlations are in fact 2D at this temperature. The $J_b$ interaction, coupling the two inequivalent sites, probably plays the leading role in these correlations. 

There is multiple evidence for a change of regime below $T^*\approx0.7$~K, which we associate with a crossover from 2D to 3D short range correlations, and which is responsible for a significant slowing down of the fluctuations. The most prominent features are the reduction of $\xi_c$ and the freezing of the incommensurability at low temperatures. Both can be understood from a rearrangement of the 2D short range ordered clusters to satisfy the coupling along the $a$-axis. Large number of moments must therefore flip, partially reducing $\xi_c$ in the process and also affecting the incommensurability along the $c$-axis. This is supported by the increase of the correlation length along the $a$ and $b$ axes at low temperatures (Fig.~\refsub[d-e]{RITA/TASP_RITA_dataComparison}). 

This crossover is also apparent in susceptibility measurements. For fields applied along the $a$-axis, Hayes \textit{et al.} observed a difference in the field cooled and zero-field cooled (ZFC) susceptibility data below $T=0.7$~K.\cite{J.Hayes2012} They also report long relaxation times (several hours) of the magnetization at $T=0.5$~K when a field is applied to a sample from ZFC conditions. Furthermore, our AC susceptibility measurements indicate a freezing temperature of $T_F\approx0.8$~K, supporting the presence of slow fluctuations in the 3D regime of \sdo. The rearrangement of 2D spin clusters into 3D correlated structures at these low temperatures can be expected to be a slow process and trap defects on a long time scale, precluding long range order. The decoupling of the two sublattices presented earlier increases the degeneracy of the system, also reducing the tendency to long range order. 

\subsection{Domain walls in zig-zag chains}
At very low temperatures, the system is in a regime of slow fluctuations, as evidenced by the AC susceptibility. For a time scale shorter than the fluctuations, the system can be described by different magnetic domains separated by domain walls. The dynamics of these domain walls can help to understand how the system avoids long range order. We will discuss first in detail the 1D case and then consider the 2D and 3D cases. We will limit ourselves to the case of $J_2/J_1>0.5$ with $J_1$ and $J_2$ being antiferromagnetic featuring a double N\'eel ground state at $T=0$~K. A very similar discussion was made for ferromagnetic $J_1$ by Redner $\&$ Krapivsky.\cite{Redner1998} 

\begin{figure}[!htb]
\includegraphics[scale=0.55]{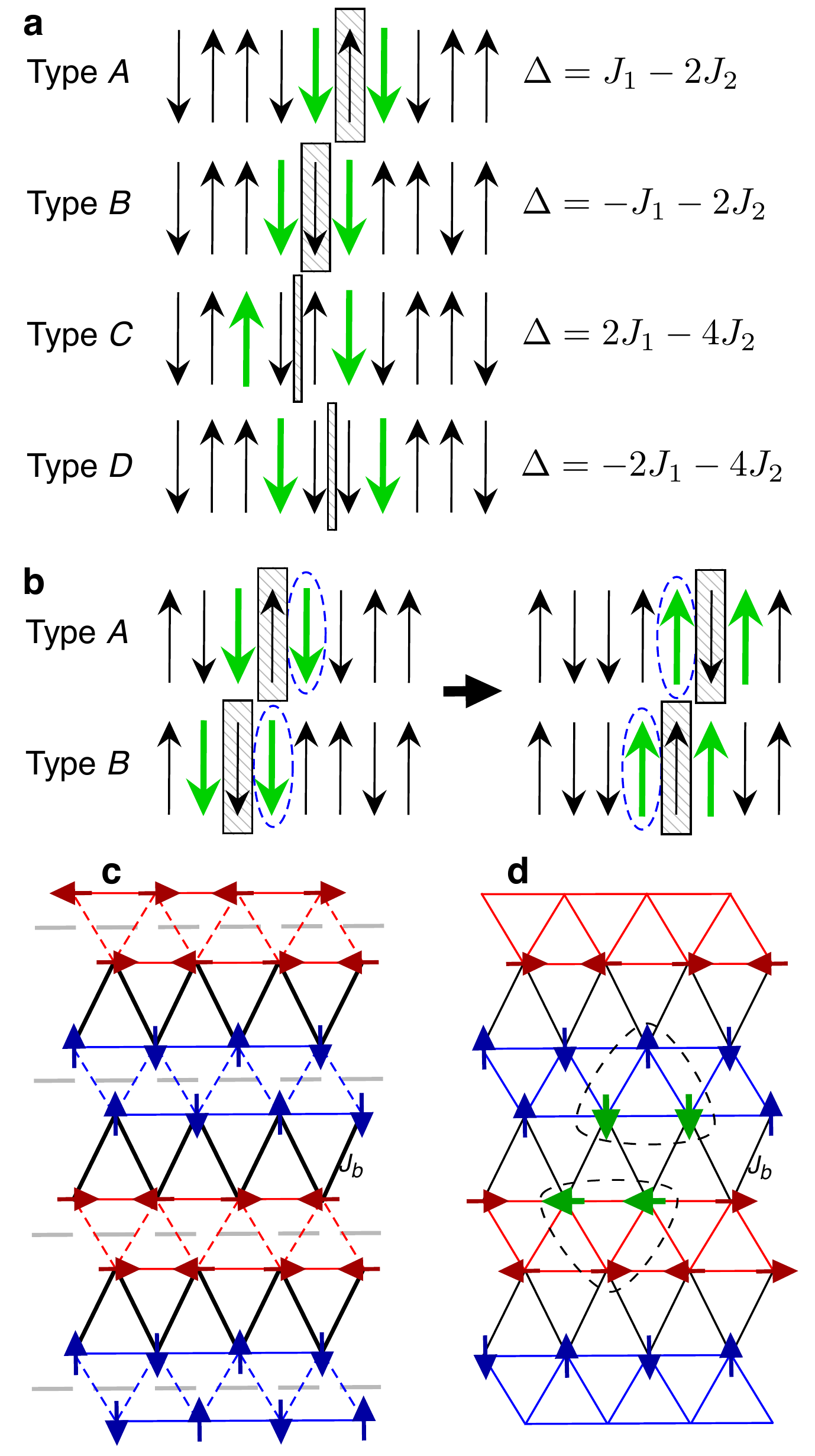}
\caption{(Color online) (a) Four different types of domain walls for the double N\'eel state of the 1D ANNNI model. The domain walls are schematized by the box and the bold green arrows represent \textit{free}-spins (see text). (b) Propagation of $A$ and $B$ domain walls by single spin flip. The circled spin is flipped. (c) Example of the configuration in the $bc$ plane without defects. The grey dashed lines represent the decoupling of the spins on one zig-zag chain due to the double N\'eel state. The resulting 1D stripes are non-frustrated zig-zag chains formed from both inequivalent sites represented in red and blue. (d) Example of configuration in the $bc$ plane in the presence of a pair of $A$ defects. The defects are circled by dashed lines and \textit{free}-spins are represented by bold green arrows. }
\label{domainwalls}
\end{figure}

The zig-zag chain with the double N\'eel ground state has four degenerate states obtained by shifting the \uudd\ pattern. For simplicity, the moment patterns are presented in the 1D ANNNI model description, i.e. on a linear chain with nearest-neighbour interaction $J_1$ and next-nearest-neighbour $J_2$. Four types of domain walls (or defects) can be formed, shown in Fig.~\refsub[a]{domainwalls} along with their associated energy cost relative to the ordered ground state. Defects of type $A$ and $B$ are the basic elements because $C$ and $D$ defects are simply the combination of two $A$ and $B$ defects respectively. On both sides of these defects, there are always two \textit{free}-spins, depicted by green bold arrows on Fig.~\refsub[a]{domainwalls}. These \textit{free}-spins can flip without changing the energy since the interaction energy at their position cancels out exactly. A single flip of a \textit{free}-spin results in a shift of type $A$ and $B$ defects, as shown on Fig.~\refsub[b]{domainwalls}. These defects can therefore propagate freely along the chain and their motion resulting from multiple spin-flip processes is related to the activation energy observed in the AC susceptibility. 

When two $A$ defects meet, they form a type $C$ defect without energy cost, which can split back to two $A$ defects. When two $B$ defects meet, they form a type $D$ defect without energy cost, which can split back to $B$ defects but can also lower its energy by splitting into two $A$ defects. These two processes are summarized by (1) $A+A \rightarrow C \rightarrow A+A$ and (2) $B+B \rightarrow D \rightarrow A+A$ where the energy reduction decay is favoured. When one $A$ defect collides with one $B$ defect, they annihilate and create a defect-free state ((3) $A+B \rightarrow 0$). Processes 2 and 3 thus represent the main decay channels of the $B$ defects. However, only process 3 can remove $A$ defects and it can only happen in the presence of $B$ defects which have a higher decay rate. More complex processes are needed to describe completely the decay of $A$ defects. It can happen through the 3-body process $A+A+A \rightarrow B$, which is energetically favoured for $J_2/J_1>1$. However for the regime with $0.5<J_2/J_1<1$, 4-body processes have to be taken into account.\cite{Cheon2001} In such a case the decay rate of $A$ defects is significantly lower and the formation of large domains can be very long. The ratio $J_2/J_1=0.63$ of \sdo\ places it in this regime. 

We now consider the behaviour of these defects in 2D and more precisely in the $bc$ plane assuming that $J_b$ is the interaction responsible for the correlations along the $b$-axis and that both inequivalent sites have the double N\'eel ground state. The topology of the interactions leads to a perfect decoupling between nearest-neighbours in the double N\'eel state. Therefore, the system in its ground state can be described as independent 1D ordered stripes (Fig.~\refsub[c]{domainwalls}). The presence of a single A or B defect on a chain is unstable because all $J_b$ interactions on one of the chain segments will be unsatisfied. It can be stabilized by adding the appropriate defect on the neighbouring zigzag chain to have $J_b$ interactions satisfied on both sides of the defect pair (Fig.~\refsub[d]{domainwalls}). A single spin flip process allows this pair to propagate freely along the direction of the zig-zag chain. These defects are strongly bound and experience a confinement potential $E=NJ_b$ for a separation of $N$ sites. The pair can be formed of type $A$, $B$ or both. The decay processes are expected to be the same as for the purely 1D case. There is no binding between pairs of defects and therefore no proper 2D domain walls but only defect pairs propagating freely along the $c$-axis.

When introducing the $J_a$ interaction along the $a$-axis, the system is 3D and the defect pairs are bound creating 2D-like domain walls in the $ac$ plane. It is possible to move the domain walls around freely by single spin flip processes. However, this implies many flips that need to occur in the appropriate sequence for a significant displacement. Therefore, the dynamics are expected to be significantly slower in a 3D correlated regime. This scenario is consistent with the dimensionality crossover discussed in the previous section. The system is effectively 2D above $T^*$ and defect pairs are free to propagate. Therefore, the fluctuations along the zig-zag chains are not significantly affected by 2D correlations and the 1D ANNNI model is a good description of the system. Below $T^*$, the 3D correlations lead to the formation of extended domain walls that hinders the propagation of defects along the chains and slow down the dynamics. This leads to the signature of spin freezing in AC susceptibility around $T^*$, as seen in Fig.~\ref{XT}. 

\section{Conclusions}

Our results indicate that the low temperature state of \sdo\ originates from a complex interplay of a dimensionality crossover and competing magnetic interactions. In the "high temperature" regime above $T^*\approx0.7~K$, the system is well described by the 1D ANNNI model. As the temperature is lowered, the spin fluctuation rate reduces and long-lived magnetic domains start to be stabilized. Domain walls are free to propagate but two types of elementary domain walls have different decay rates. The type $A$ domain wall is expected to decay very slowly for $0.5<J_2/J_1<1$. As the temperature is lowered below $T^*$, weaker interchain interactions such as the dipolar interactions start to be significant and the system can not be considered one dimensional anymore. This dimensionality crossover leads to competition between the domain walls decay processes and the interchain interactions, precluding long range ordering on experimental time scales in \sdo.

This compound shares many similarities with the spin ices Dy$_2$Ti$_2$O$_7$ and Ho$_2$Ti$_2$O$_7$. In these systems, the short range correlations are dictated by the ice rules that lead to a macroscopical degeneracy. At low temperatures, these spin liquids have slow fluctuations that are dominated by defects which are described as magnetic monopoles.\cite{Castelnovo2008} In \sdo, the short range correlations are described by the 1D ANNNI model and each chain has four degenerate configurations of the double N\'eel state. At low temperatures, the fluctuations are slow as seen in AC susceptibility and can be described by defects in the chains. We have evidenced that the dynamics of defects is of great importance in the low temperature physics in \sdo, a general principle which is also relevant in the spin ices.

\section*{Acknowledgements}
The authors are thankful to M. Sigrist, M. Gingras, O.~A. Petrenko, and D.~L. Quintero-Castro for fruitful discussions; M. Bartkowiak and M. Zolliker for the assistance with the dilution refrigerator experiments at SINQ. 
The crystal and magnetic structure figures have been generated with the SpinW package for Matlab.\cite{Toth2015}
This research received support from the Swiss National Foundation (SNF Grant No. 138018), the Natural Sciences and Engineering Research Council of Canada (Canada), the Fonds Qu\'eb\'ecois sur la Nature et les Technologies (Qu\'ebec) and the Canada Research Chair Foundation (Canada). 

\appendix*
\section{Calculation of the Zeeman splitting tensor}
\label{appendixGtensor}
For a doublet ground state, it is possible to write the Zeeman energy Hamiltonian in terms of an effective spin $S=\frac{1}{2}$. The original Hamiltonian is defined by
$\mathcal{H}_\text{Zeeman}=\sum_{\alpha} H_\alpha  \hat{M}_\alpha$
where the index $\alpha$ indicates the axes $x$, $y$ and $z$, $H_\alpha$ is the magnetic field and $\hat{M}_\alpha$ is the magnetization operator. The magnetization operator is defined as $\hat{\bf{M}}=\hat{\bf{L}}+2\hat{\bf{S}}$, where $\hat{\bf{L}}$ and $\hat{\bf{S}}$ are the orbital and spin operators respectively. The Hamiltonian can be rewritten for an effective spin $S=\frac{1}{2}$ as
$\mathcal{H}_\text{Zeeman}=\mu_B \sum_{\alpha \beta} H_\alpha g_{\alpha \beta} \hat{S}_\beta$
where $\hat{S}_\alpha$ is the effective spin-$\frac{1}{2}$ operator and $g_{\alpha \beta}$ forms the $\bf{g}$-tensor. Assuming that the doublet wavefunctions are $\ket{+}$ and $\ket{-}$, the elements of the $\bf{g}$-tensor can be evaluated from the matrix elements of $\hat{\bf{M}}$:
\begin{equation}
\begin{aligned}
g_{\alpha x} &= 2 \text{Re} \bra{+} \hat{M}_\alpha \ket{-}/ \mu_B, \\
g_{\alpha y} &= -2 \text{Im} \bra{+} \hat{M}_\alpha \ket{-}/ \mu_B, \\
g_{\alpha z} &= 2  \bra{+} \hat{M}_\alpha \ket{+}/ \mu_B.
\end{aligned}
\end{equation}
The $g$-factors along the principal axes are obtained from the matrix $\bf{G}=\bf{g} \cdot  \bf{g}^T$. The square-root of the eigenvalues of $\bf{G}$ are the $g$-factors along the principal axes, which are themselves defined by the eigenvectors.\cite{Weil2006,Malkin2015} 
\newline

In the presence of a mirror plane perpendicular to $z$ at the magnetic ion site, the matrix $\bf{G}$ simplifies to
\begin{equation}
\bf{G}=\begin{bmatrix}    G_{xx} &  G_{xy}   & 0 \\  G_{xy}  &  G_{yy}   & 0 \\    0 &   0 & G_{zz} \end{bmatrix}.
\end{equation}
In that case, one of the principal axis points along $z$ while the two others are in the $xy$-plane. This is the case for the magnetic ions in \sdo\ with a mirror plane perpendicular to the $c$-axis.
\newline

To evaluate the $\bf{g}$-tensor of the Dy$^{3+}$ ions in \sdo, the matrix elements of the magnetization operator $\hat{\bf{M}}$ have to be calculated for the doublet ground state. The wavefunctions obtained by the \textit{multiX} software were used for this calculation. The \textit{multiX} software \cite{Uldry2012} calculates the single-ion energy levels in the presence of crystalline electric fields. Slater determinants of the electronic configuration are used as the wavefunction basis. The matrix elements of $\hat{\bf{M}}$ have been calculated from this basis by using the Slater-Condon rules for the doublet ground state of the two inequivalent Dy$^{3+}$ sites. The $\bf{g}$-tensor, $g$-factors and principal axes were then calculated according to equations presented here. 


\end{document}